\documentclass[superscriptaddress,
reprint,
nobibnotes,
 amsmath,amssymb,
 aps,
prb,
floatfix,
]{revtex4-2}

\usepackage[utf8]{inputenc}
\usepackage{graphicx}
\usepackage{dcolumn}
\usepackage{bm}
\usepackage{xcolor}
\usepackage{microtype}
\usepackage[
    separate-uncertainty = true,
    locale=US,
    ]{siunitx}
\usepackage{cleveref}

\begin{document}

\title{Controlling bubble and skyrmion lattice order and dynamics via stripe domain engineering in ferrimagnetic Fe/Gd multilayers}

\author{Tim Titze}
\affiliation{I.\,Physikalisches Institut, Universit\"at G\"ottingen, 37077 G\"ottingen, Germany\looseness=-1}
\author{Sabri Koraltan}
\affiliation{Institute of Applied Physics, TU Wien, Vienna, A-1040, Austria}
\affiliation{Physics of Functional Materials, Faculty of Physics, University of Vienna, Vienna, Austria\looseness=-1}

\author{Timo Schmidt}
\affiliation{Institute of Physics, University of Augsburg, 86135 Augsburg, Germany\looseness=-1}
\author{Mailin Matthies}
\affiliation{I.\,Physikalisches Institut, Universit\"at G\"ottingen, 37077 G\"ottingen, Germany\looseness=-1}
\author{Amalio~Fern\'{a}ndez-Pacheco}
\affiliation{Institute of Applied Physics, TU Wien, Vienna, A-1040, Austria}
\author{Dieter Suess}
\affiliation{Physics of Functional Materials, Faculty of Physics, University of Vienna, Vienna, Austria\looseness=-1}
\author{Manfred Albrecht}
\affiliation{Institute of Physics, University of Augsburg, 86135 Augsburg, Germany\looseness=-1}
\author{Stefan Mathias}
\affiliation{I.\,Physikalisches Institut, Universit\"at G\"ottingen, 37077 G\"ottingen, Germany\looseness=-1}
\affiliation{International Center for Advanced Studies of Energy Conversion (ICASEC), Universit\"at G\"ottingen, 37077 G\"ottingen, Germany}
\author{Daniel Steil}
\affiliation{I.\,Physikalisches Institut, Universit\"at G\"ottingen, 37077 G\"ottingen, Germany\looseness=-1}

\begin{abstract}
Ferrimagnetic Fe/Gd multilayers host maze-like stripe domains that transform into a disordered bubble/skyrmion lattice under out-of-plane magnetic fields at ambient temperature. Femtosecond magneto-optics distinguishes these textures via their distinct coherent breathing dynamics. Crucially, applying a brief in-plane ``set'' magnetic field to the stripe state enhances both frequency and amplitude of the bubble/skyrmion lattice breathing mode. Lorentz transmission electron microscopy, magnetic force microscopy, and micromagnetic simulations reveal that this enhancement arises from field-aligned stripes nucleating a dense, near-hexagonal bubble/skyrmion lattice upon out-of-plane field application, with strong indications for a pure skyrmion lattice. Thus, modifying the initial domain configuration by in-plane fields enables precise control of coherent magnetization dynamics on picosecond to nanosecond timescales and potentially even of topology.
\end{abstract}

\maketitle

\section{Introduction}
Magnetic skyrmions, first observed in 2009 by Mühlbauer et al.~\cite{Muehlbauer2009}, have attracted significant interest due to their tremendous potential in novel spintronic and magnonic devices~\cite{Nagaosa2013,Finocchio2016,Fert2017, Grollier2020, Yu2021}. One key reason is their intricate spin configuration, which gives rise to topological properties resulting in enhanced stability against perturbations. Magnetic skyrmions are commonly stabilized by the Dzyaloshinskii-Moriya interaction (DMI), either of interfacial~\cite{Heinze2011,Chen2015,Woo2016} or bulk type~\cite{Muehlbauer2009,Muenzer2010,Yu2010}. However, they can also exist in multilayer thin films without DMI, where they are stabilized by the competition between dipolar interactions and perpendicular magnetic anisotropy~\cite{Yu2012,Lee2016FeGd,Desautels2019,Zhang2020c,Montoya2017a,Heigl2021,Hassan2024}. In such dipolar-stabilized systems no preferred chirality is imposed, allowing the coexistence of skyrmions of both helicities (type-1 bubbles, topological charge $Q=-1$) and topologically trivial type-2 bubbles (topological charge $Q=0$). The ordering and density of these objects are of great interest for fundamental studies and future technological applications, e.g., in the context of magnonic crystals~\cite{Chumak2017, Lonsky2020, Zhedong2021}. Achieving active control over their formation is therefore a key step toward unlocking the full potential of easy-to-grow, easily tunable multilayer thin films for skyrmion-based devices.   

One particular system hosting mixed lattices of bubbles and skyrmions at ambient temperature are metallic ferrimagnetic Fe/Gd multilayers, which exhibit maze-like magnetic stripe domains transforming into a mixed bubble and skyrmion (BSK) lattice within a certain range of applied out-of-plane magnetic fields~\cite{Montoya2017a,Heigl2021,Titze2024a}. 

The different magnetic field–dependent spin textures in this system can, for example, be identified by observing the distinct coherent responses of the spin system following a femtosecond optical stimulus, a technique sucessfully applied previously for other systems as well~\cite{Ogawa2015, Padmanabhan2019,Kalin2022}. This way, stripe domain phases as well as BSK lattice states can be unequivocally distinguished in the time-domain without resorting to magnetic imaging methods~\cite{Titze2024a}. Furthermore, strong optical excitation of the inherent magnetic spin textures allows for their nucleation or annihilation~\cite{Titze2024a,Titze2025}, and even coherent control of the dynamics of BSK lattice states is possible using a double-pump excitation scheme~\cite{Titze2024b}. 

Here, we showcase how controlling the initial stripe domain magnetic textures in Fe/Gd multilayers determines the density and ordering of the resulting cylindrical spin objects, achieving a highly dense, close to ideally ordered hexagonal BSK lattice. This effect is achieved by applying an in-plane magnetic set field to the sample prior to increasing the stabilizing out-of-plane magnetic field, which leads to a strongly enhanced coherent response in femtosecond magnetooptics and a large density of spin objects as observed in Lorentz transmission electron microscopy (LTEM), magnetic force microscopy (MFM), and supported by micromagnetic simulations. LTEM and micromagnetic simulations further indicate that this lattice is predominantly of skyrmion type, inferring the creation of a nearly perfect skyrmion lattice. 

\section{Results}
\subsection{Ground state magnetic system properties}
\begin{figure}[t]
     \centering
     \includegraphics[width=1\columnwidth]{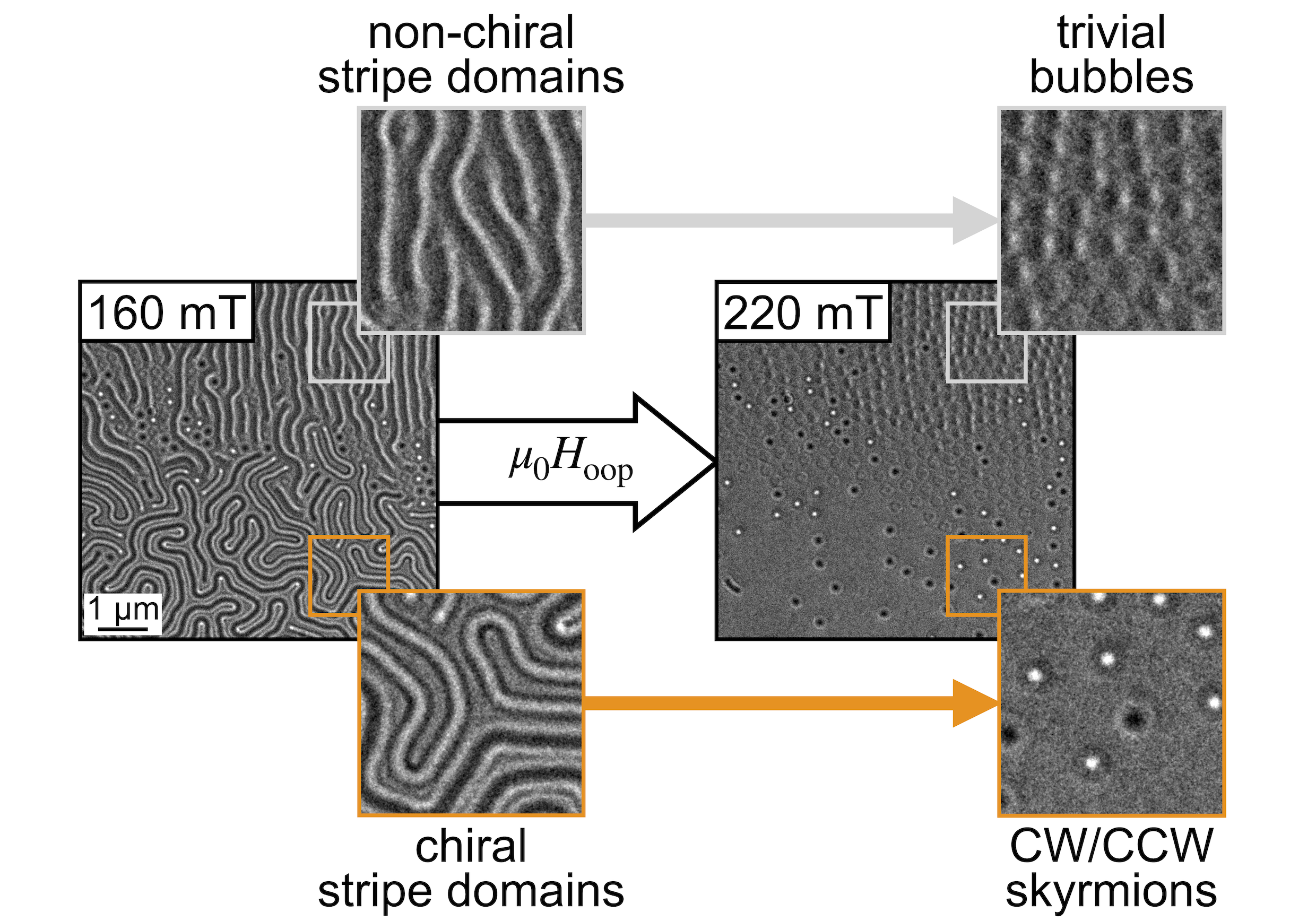}
     \caption{\textbf{Ground state magnetic spin textures hosted by the [Fe($0.35$\,nm)/Gd($0.40$\,nm)]$_{120}$ multilayer film for LTEM studies.} At low magnetic fields, the system exhibits two distinct types of stripe domains: chiral and non-chiral stripe domains. At high out-of-plane magnetic fields, the chiral stripe domains have been transformed into clockwise (CW) and counterclockwise (CCW) skyrmions (orange), while topologically trivial bubbles have emerged from the non-chiral ones (gray).}
     \label{fig:Groundstate}
\end{figure}

Shortly summarizing our previous studies, see Refs.~\cite{Titze2024a, Titze2025}, the studied Fe/Gd multilayers with composition [Fe($0.35$\,nm)/Gd($0.40$\,nm)]$_{160}$ show different kinds of spin textures under out-of-plane magnetic fields. Figure~\ref{fig:Groundstate} displays the ground state magnetic spin textures stabilized at two exemplary magnetic fields ($\mu_0H_{\mathrm{oop}}=160$\,mT and $\mu_0H_{\mathrm{oop}}=220$\,mT). For the lower magnetic field, we find two distinct types of stripe domains, namely chiral and non-chiral stripe domains classified by the rotation sense of the magnetic moments across the domain walls~\cite{Yu2012}. Chiral stripe domains exhibit a full rotation of magnetic moments, either clockwise or counterclockwise, across both domain walls. In contrast, non-chiral stripe domains display opposite rotation senses at the two domain walls, i.e., they lack a defined chirality.  
These can be well distinguished in LTEM since non-chiral domain walls show a white-black contrast, while chiral domain walls exhibit a pronounced black-white-black or white-black-white contrast depending on their chirality, i.e., left- or right-handed rotation sense. When increasing the out-of-plane magnetic field, these stripe domains transform into cylindrical spin objects, namely topologically trivial bubbles as well as clockwise (CW) and counterclockwise (CCW) Bloch-type skyrmions. However, bubbles only emerge from non-chiral stripe domains while skyrmions nucleate from chiral stripe domains with the corresponding chirality, meaning the chirality is preserved during the nucleation process.

\subsection{Magnetooptical studies of coherent spin dynamics}
\begin{figure}[t]
     \centering
     \includegraphics[width=1\columnwidth]{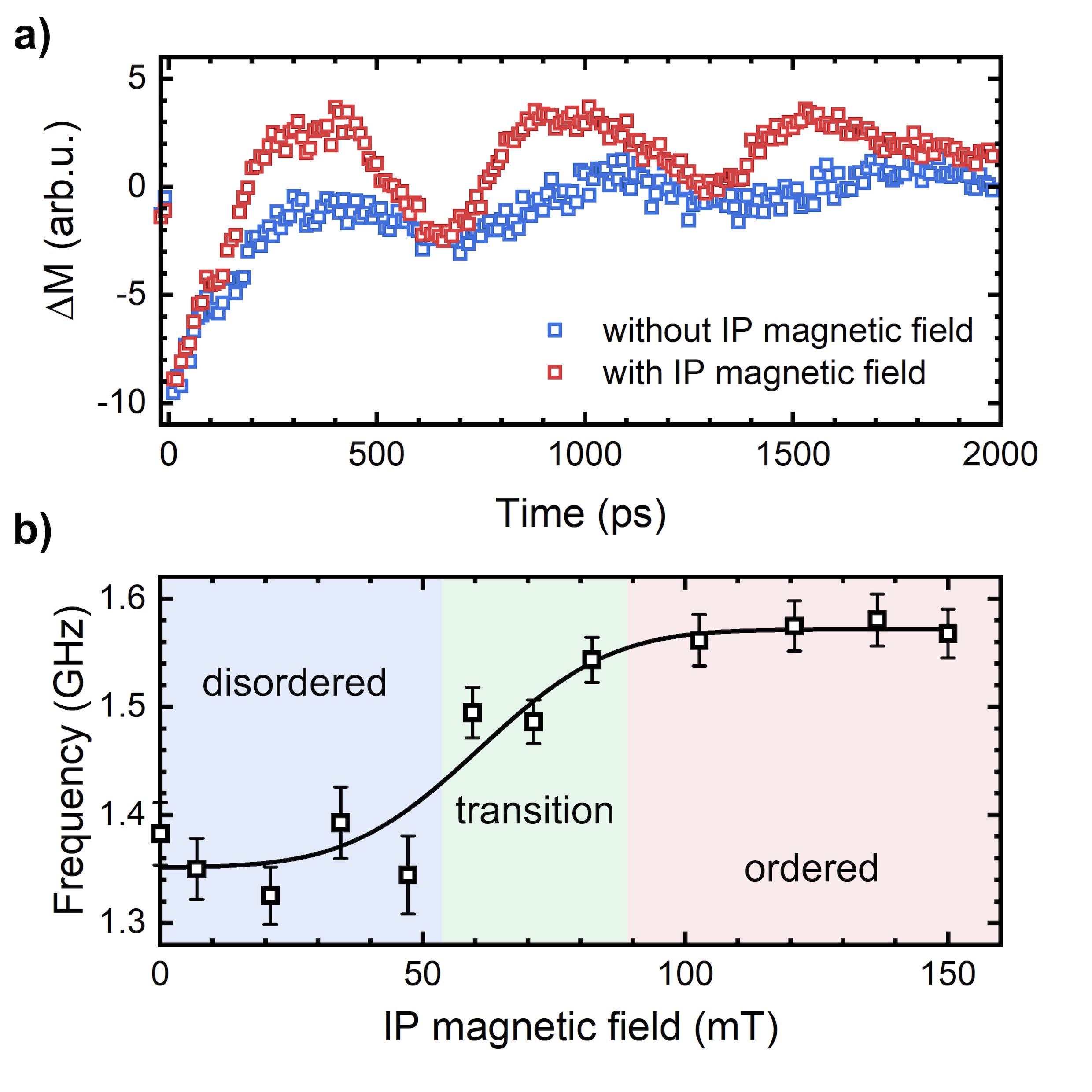}
     \caption{\textbf{The effect of initial in-plane magnetic set fields on the coherent magnetization dynamics.} \textbf{a)}~Laser-induced magnetization dynamics using an out-of-plane magnetic field $\mu_0H_{\mathrm{oop}}=196$\,mT. In the case of the red squares, an in-plane magnetic set field $\mu_0H_{\mathrm{ip}}=151$\,mT was switched on prior to the measurement to align the demagnetized stripe domain state. \textbf{b)}~Fourier analysis of the breathing mode frequency for various in-plane magnetic set fields which were applied prior to the constant out-of-plane magnetic field $\mu_0H_{\mathrm{oop}}=196$\,mT.}
     \label{fig:Data}
\end{figure}

Following our recent works on detection of BSK lattices via femtosecond magnetooptics, Figure~\ref{fig:Data}a depicts the laser-induced magnetization dynamics of the Fe/Gd multilayer system with composition [Fe($0.35$\,nm)/Gd($0.40$\,nm)]$_{160}$ in the weak perturbation limit, $F=0.7$\,mJ/cm$^2$. The data shown was recorded using an out-of-plane magnetic field $\mu_0H_{\mathrm{oop}}=196$\,mT, i.e., in the BSK lattice state. Here, the blue transient shows the characteristic presence of the coherent BSK breathing mode~\cite{Titze2024a,Titze2024b,Titze2025} under application of only the out-of-plane magnetic field. Interestingly, application of an in-plane magnetic field $\mu_0H_{\mathrm{ip}}=151$\,mT prior to the measurement significantly alters the observed dynamics, as depicted by the red transient. More specifically, we observe both an increase in amplitude and frequency of the breathing mode. Fourier analysis and fitting using a Gaussian error function reveals a frequency upshift of $\Delta f\approx 0.2$\,GHz while increasing the in-plane magnetic field, see Fig.~\ref{fig:Data}b. In addition, we obtain a threshold in-plane magnetic field $\mu_0H_{\mathrm{th,ip}}\approx 55$\,mT that is required to drive the observed change in dynamics. This transition appears to be continuous and in-plane magnetic fields $\mu_0H_{\mathrm{ip}}\geq 100$\,mT are required to achieve a full transformation.
\begin{figure*}[t]
     \centering
     \includegraphics[width=1\textwidth]{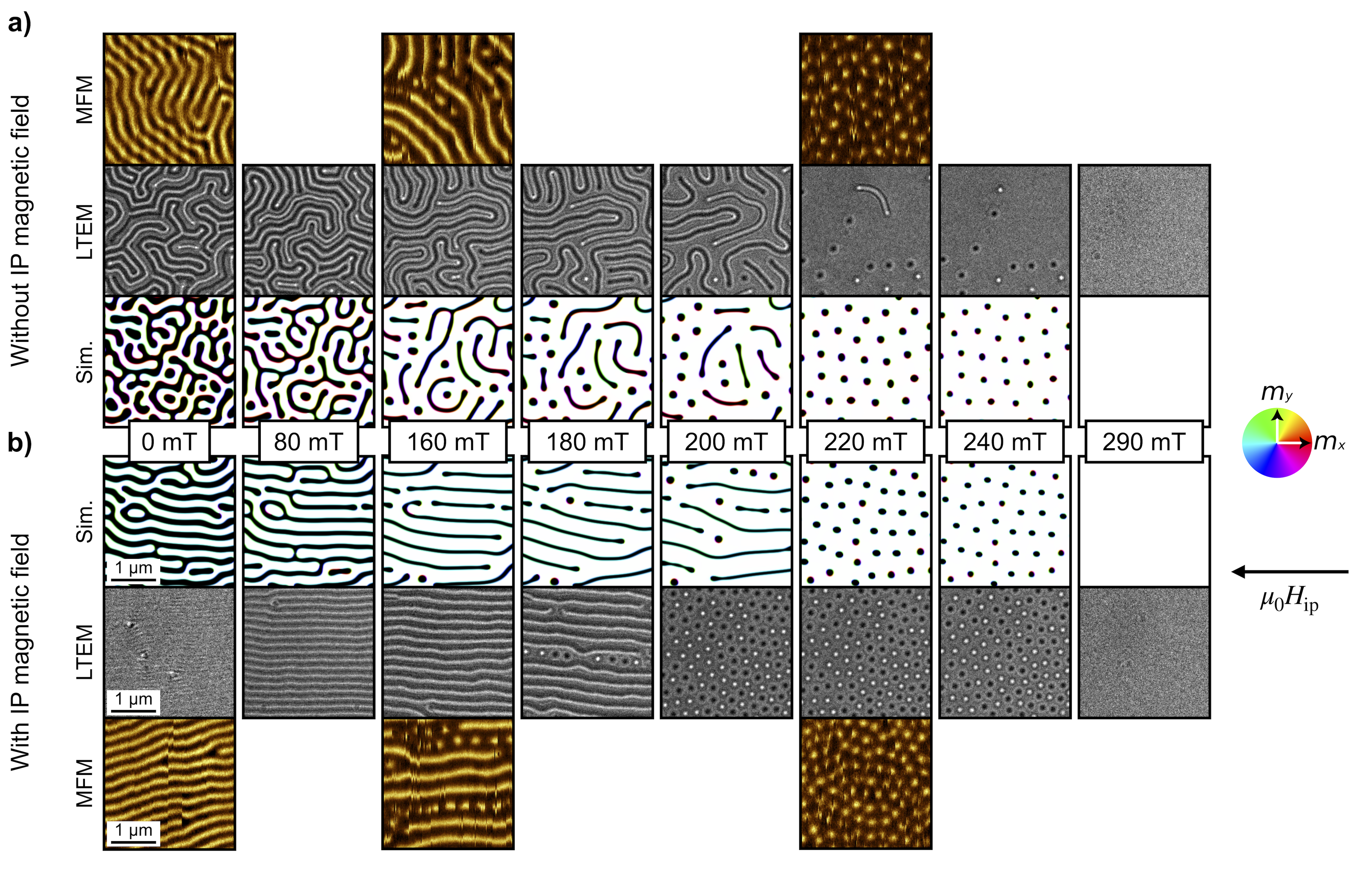}
     \caption{\textbf{The effect of in-plane magnetic set fields on the magnetic spin textures.} \textbf{a)}~Magnetic spin textures obtained from micromagnetic simulations, \textsc{LTEM}, and \textsc{MFM} considering various out-of-plane magnetic fields as given in the center of the image. \textbf{b)}~An in-plane magnetic set field was applied prior to increasing the out-of-plane magnetic field to align the stripe domain state. For the region shown in the \textsc{LTEM} images, which is approximately the same for all images, membrane buckling effects are minimized. The \textsc{LTEM} measurements were carried out on a [Fe($0.35$\,nm)/Gd($0.40$\,nm)]$_{120}$ sample, while the \textsc{MFM} measurements and micromagnetic simulations were performed on the [Fe($0.35$\,nm)/Gd($0.40$\,nm)]$_{160}$ sample which was used for the optical experiments.}
     \label{fig:Simulation}
\end{figure*}

This modification occurs solely due to the application of an in-plane magnetic field before the actual time-resolved measurements, suggesting that it arises from changes in the underlying static magnetic spin textures of the system. Therefore, we now compare the two different pathways induced by the initial magnetic domain configuration, which can be tailored by applying in-plane magnetic set fields. The resulting spin textures in out-of-plane magnetic fields are analyzed by LTEM, MFM, and micromagnetic simulations.

\subsection{Imaging of magnetic spin textures}
Figure~\ref{fig:Simulation}a shows the magnetic spin textures obtained from LTEM, MFM, and micromagnetic simulations by applying only an out-of-plane magnetic field. Note that the LTEM images were obtained from a [Fe($0.35$\,nm)/Gd($0.40$\,nm)]$_{120}$ multilayer sample on a $30$\,nm thick Si$_3$N$_4$ membrane, which has somewhat different magnetic properties than the rigid thin film in MOKE experiments due to membrane buckling effects (see Appendix). In contrast, the MFM data was taken from a piece of the [Fe($0.35$\,nm)/Gd($0.40$\,nm)]$_{160}$ sample studied in the time-resolved measurements. Therefore, LTEM data, which allows to clearly distinguish chiral and non-chiral stripe domains, as well as bubbles and skyrmions, is used for qualitative understanding of spin texture changes due to additional in-plane fields, whereas MFM data is used for quantitative understanding of field-induced changes. By increasing the out-of-plane magnetic field from $\mu_0 H_\textrm{oop}=0$\,mT up to $\mu_0 H_\textrm{oop}=290$\,mT, the magnetic spin textures change from a maze-like stripe domain state into a mixed state of stripe domains and cylindrical spin objects, and finally into a disordered BSK lattice, before the single domain state is formed. In more detail, micromagnetic simulations predict a mixed lattice of bubbles and skyrmions with a density of $4.4$\,{\textmu}m$^{-2}$ at $\mu_0 H_\textrm{oop}=220$\,mT,  lower than the density of cylindrical spin objects observed for the rigid thin film sample in MFM, which is $7.5$\,{\textmu}m$^{-2}$. LTEM data, in contrast shows only a low number, respectively density of purely skyrmionic objects in the depicted field of view (FOV). We note that the density and type of spin objects in LTEM strongly depends on the aforementioned membrane buckling, which is expected to be minimized in the given FOV. Thereby, considering magnetic properties, this FOV should correspond best to the thin film on the rigid Si substrate used in MFM and optical measurements, even though the total number of skyrmions is low. In regions that exhibit stronger membrane buckling, both bubbles and skyrmions are present as shown in Figs.~\ref{fig:Groundstate} and~\ref{fig:Membrane}. 

Figure~\ref{fig:Simulation}b now shows snapshots of the magnetic spin textures for the same out-of-plane magnetic fields, but before increasing the out-of-plane magnetic field from zero to saturation, an in-plane magnetic set field was temporarily applied. The effect of an applied in-plane field of $\mu_0 H_\textrm{ip}=50$\,mT at zero out-of-plane magnetic field in the micromagnetic simulation is an alignment of the initial maze-like stripe domains, now forming Bloch-type domain walls parallel to the in-plane field. Turning this field off preserves the parallel domain wall alignment. Gradually increasing the external out-of-plane magnetic field in the simulation to $\mu_0 H_\textrm{oop}=220$\,mT, once again transforms oriented stripes into a mixed BSK lattice with an initially large number of bubbles, see also Fig.~\ref{fig:SimSingleSkyrmion}. Intriguingly, skyrmions nucleate despite the stripe domains lacking chirality. In addition, the resulting lattice exhibits an about 20\% higher density ($5.2$\,{\textmu}m$^{-2}$) at $\mu_0 H_\textrm{oop}=220$\,mT than in the previous case ($4.4$\,{\textmu}m$^{-2}$) and an ordering, which is now closer to an ideal hexagonal lattice structure. Further increase of the magnetic field yields a single skyrmion lattice (see also Fig.~\ref{fig:SimSingleSkyrmion}), i.e., the initially created bubbles appear to be extremely unstable. A similar transitional bubble state in a stripe domain to skyrmion transformation under the influence of an additional in-plane magnetic field was recently discussed in Ref.~\cite{Jefremovas2025} for Néel-type skyrmions stabilized by DMI, also leading to a single skyrmion lattice with enhanced density. 

The findings in the simulation are in very good agreement with the results obtained from LTEM, which also showcase the alignment of the domain walls by the in-plane field. In LTEM, application of an aligning in-plane field is achieved by tilting the sample by $30^\circ$ in an applied out-of-plane field prior to the application of the actual out of plane field at zero tilt angle. However, in contrast to the micromagnetic simulations, non-chiral stripe domains seem to directly transform into a single skyrmion lattice upon increasing the out-of-plane magnetic field. An intermediate bubble state may not be observable if the bubbles are comparably unstable as those in the simulation. Interestingly, the number of skyrmions at $\mu_0 H_\textrm{oop}=220$\,mT within the same FOV is now massively increased compared to Fig.~\ref{fig:Simulation}a, forming a dense, close to hexagonal lattice with a density of $7.7$\,{\textmu}m$^{-2}$.

The increase in the density of cylindrical spin objects is as well corroborated by the MFM measurements on the sample on the rigid substrate. Here, application of an in-plane field prior to the out-of-plane field again shows the alignment of maze-like stripe domains and the resulting density of spin objects at $\mu_0 H_\textrm{oop}=220$\,mT is $11.4$\,{\textmu}m$^{-2}$ compared to $7.5$\,{\textmu}m$^{-2}$ without application of the in-plane field, an increase of about 50\%.

In total, LTEM, MFM, and micromagnetic simulations all show that briefly applying an in-plane magnetic set field at zero out-of-plane field leads to an alignment of maze-like stripe domains and that application of a large enough out-of-plane field transforms these stripe domains into a lattice of cylindrical spin objects with increased density compared to the case without in-plane field. Furthermore, micromagnetic simulations and LTEM indicate that such a procedure may lead to the formation of a pure skyrmion lattice, as bubbles created this way seem to be very unstable in magnetic fields, see Fig.~\ref{fig:SimSingleSkyrmion} in the Appendix.

\subsection{Dynamical response to laser excitation in micromagnetic simulations}
\begin{figure}[t]
     \centering
     \includegraphics[width=1\columnwidth]{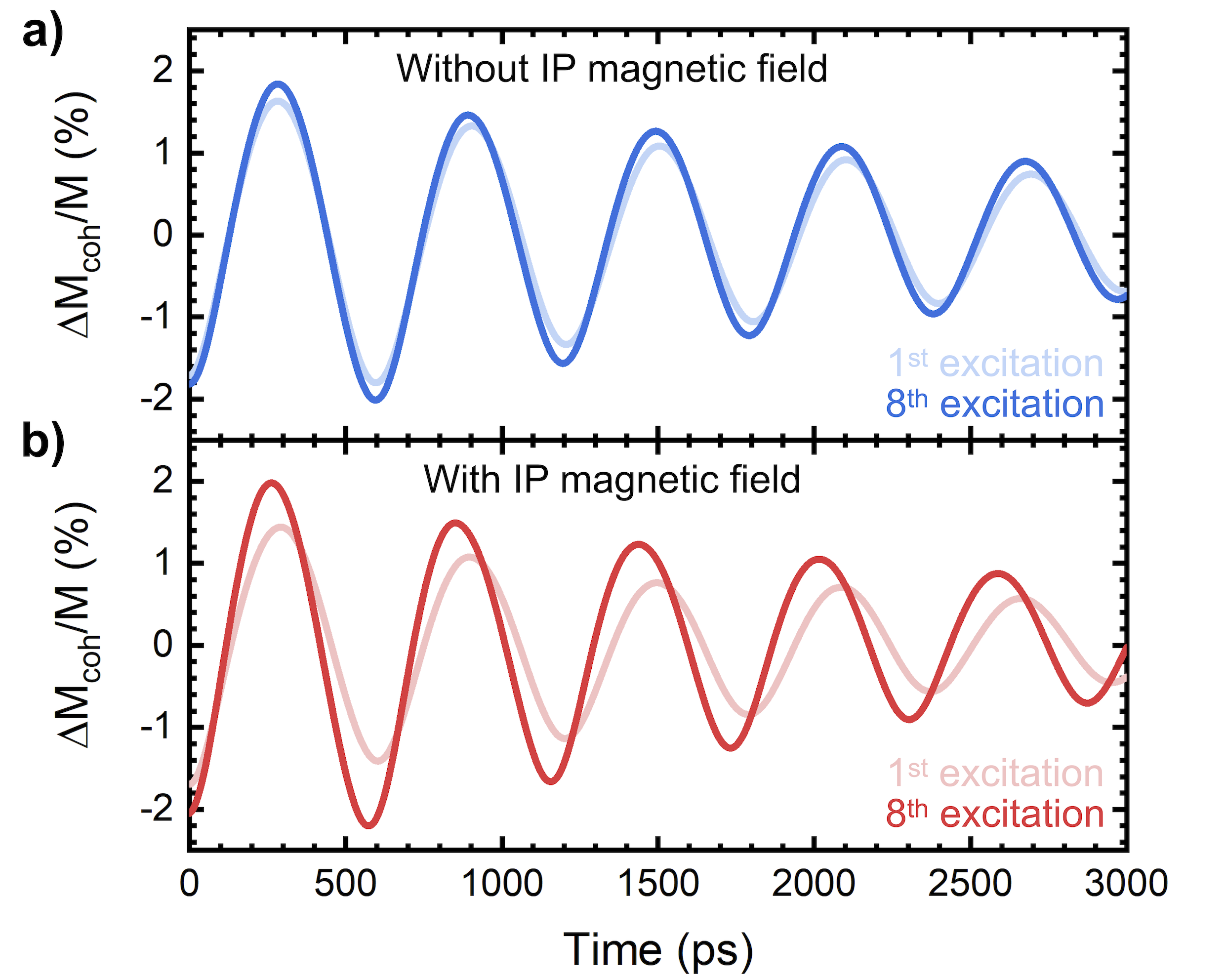}
     \caption{\textbf{Coherent magnetization dynamics obtained from micromagnetic simulations.} \textbf{a)} Coherent magnetization dynamics after 1 and 8 excitations of the initial state created without applying an in-plane magnetic field and \textbf{b)} of the initial state for which an in-plane magnetic set field has been applied prior to increasing the out-of-plane magnetic field.}
     \label{fig:Excitations}
\end{figure}
How do these observations now relate to the observed changes in both breathing mode amplitude and frequency in the MOKE experiment? To elucidate the origin of these changes, we employed dynamic micromagnetic simulations modelling the laser-induced breathing mode dynamics, following the approach of our recent studies~\cite{Titze2024a,Titze2024b}. Figures~\ref{fig:Excitations}a and \ref{fig:Excitations}b display these coherent magnetization dynamics resulting from excitation of an initial state without and with prior in-plane magnetic set field. Furthermore, we compare the dynamics obtained from a single excitation and after eight excitations, i.e., following consecutive simulated laser excitations. While the dynamics only slightly differ in case of no applied in-plane magnetic field, there is a significant change in both frequency and amplitude when an in-plane magnetic set field is used to modify the initial state. Furthermore, comparing the eighth excitation of both cases, we find an increased frequency ($\Delta f \approx 0.05$\,GHz) and amplitude when applying the in-plane magnetic field, qualitatively similar to the experiment.

\begin{figure}[t]
     \centering
     \includegraphics[width=1\columnwidth]{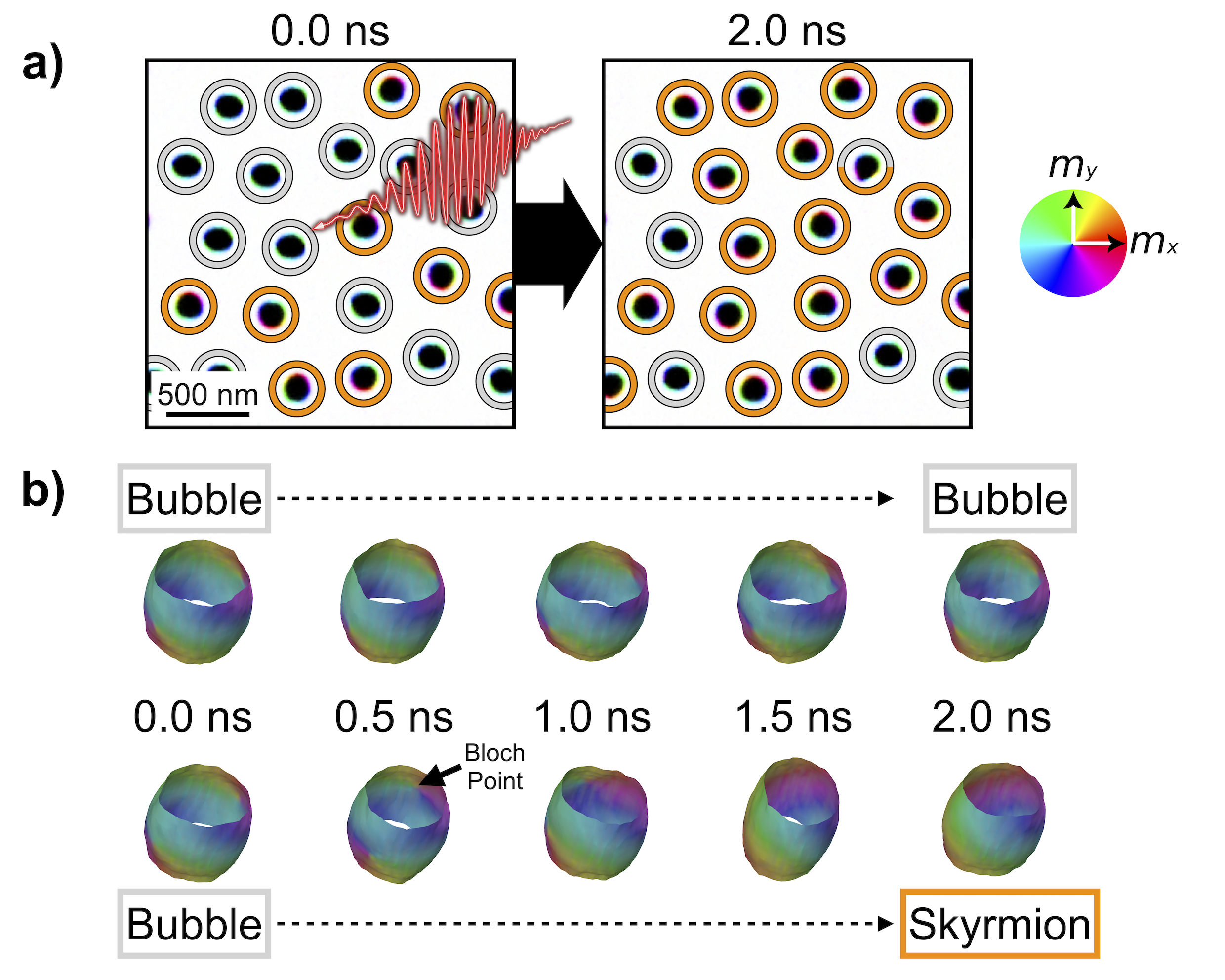}
     \caption{\textbf{Transformation of topologically trivial bubbles into non-trivial skyrmions.} \textbf{a)} The initial state in the simulation is created by briefly applying an in-plane magnetic set field of $\mu_0 H_{\mathrm{ip}}=50$\,mT before increasing the out-of-plane magnetic field to $\mu_0 H_{\mathrm{oop}}=220$\,mT. Most of the topologically trivial bubbles (highlighted by gray circles) transform into non-trivial skyrmions (orange circles) upon laser excitation. \textbf{b)} 3D representation of two topologically trivial bubbles. While the bubble in the upper panel remains unchanged, the bubble in the lower panel transforms into a non-trivial skyrmion by introducing a Bloch point at around 0.5\,ns after laser excitation. The transformation process is completed at approximately 1.5\,ns.}
     \label{fig:Transformation}
\end{figure}

The micromagnetic simulations allow us to explore the reason for these observed changes. The prior application of an in-plane magnetic set field modifies the initial state in a way that more topologically trivial bubbles are stabilized and the overall density of spin objects is increased by about 20\%, see also Fig.~\ref{fig:Simulation}. However, almost all of these bubbles are transformed into non-trivial skyrmions upon consecutive laser excitation, see Fig.~\ref{fig:Transformation}. This is the reason for the increase in frequency and amplitude of the breathing mode that we observe after eight laser excitations in the simulation, as skyrmions show a faster and more pronounced expansion and contraction of their core size than bubbles~\cite{Titze2024a}. Comparing the two cases with and without in-plane magnetic set field after eight excitations, the in-plane field treatment results in an additional increase of breathing frequency and amplitude. This can be unequivocally related to the 20\% enhanced density of spin objects as almost no bubbles remain after multiple excitations, i.e., the observed breathing mode predominantly originates from skyrmions in both cases. 
We note that the observed increase in frequency, and particularly in amplitude, is rather weak compared to the experimental data shown in Fig.~\ref{fig:Data}a. We relate this effect to the relative small increase in spin object density in the simulation in contrast to the experiment (see MFM data in Fig.~\ref{fig:Simulation}).  

Going into more detail, the aforementioned transformation of topologically trivial bubbles into non-trivial skyrmions upon laser excitation is displayed in Fig.~\ref{fig:Transformation}. We start with an initial state shown in Fig.~\ref{fig:Transformation}a, which is created by briefly applying an in-plane magnetic set field $\mu_0H_{\mathrm{ip}}=50$\,mT before increasing the out-of-plane magnetic field to $\mu_0H_{\mathrm{oop}}=220$\,mT. Laser excitation then yields the transformation of bubbles to skyrmions as evident from the snapshot at $2$\,ns after excitation. We further examine the transformation process in Fig.~\ref{fig:Transformation}b by monitoring the evolution of the 3D structure of the in-plane magnetic moments across the film thickness. The upper panel displays a bubble, which is not transformed upon laser excitation. In contrast, the lower panel shows a bubble that is transformed into a skyrmion. This transformation process is mediated by the formation of a Bloch point, a topological magnetic singularity~\cite{Kosevich1990, HierroRodriguez2020, Tejo2021, Goebel2021}, at approximately $0.5$\,ns after laser excitation and it is fully completed after $1.5$\,ns. We suggest that the laser-induced transformation of bubbles is only possible because they resemble a metastable state in this system as also pointed out by Jefremovas et al.~\cite{Jefremovas2025}. Under real experimental conditions, bubbles can therefore not be stabilized on observable timescales as highlighted by the LTEM measurements shown in Fig.~\ref{fig:Simulation}. 

\section{Discussion}
Combining the experimental results and micromagnetic simulations, we now explain the changes in the experimentally observed breathing mode amplitude and frequencies in Fig.~\ref{fig:Data}. In agreement with LTEM, MFM, and micromagnetic simulations, as well as Refs.~\cite{Zhang2018b,Jefremovas2025}, applying an in-plane magnetic field significantly increases the density of spin objects, here by about 50\% according to the MFM data. We attribute this to a unique mechanism of skyrmion nucleation from non-chiral stripe domains, leading to a pure, close to hexagonally ordered skyrmion lattice state. This skyrmion lattice emerges from an ordered, but metastable bubble lattice state after the formation of Bloch points. These Bloch points appear in the micromagnetic simulations at about $0.5$\,ns after weak laser excitation, presumably in response to the laser-induced disturbance of the system magnetization as the system recovers to a more stable state configuration. 

The higher density of spin objects gives rise to an enhanced frequency and amplitude of the breathing mode in the femtosecond experiment, as corroborated by dynamic micromagnetic simulations~\footnote{Note that these changes are considerably smaller than those observed experimentally, which is related to the smaller changes in density in the simulation compared to the MFM.}. The amplitude increases, since more spin objects contribute to the collective breathing mode. However, the higher density of spin objects also results in a reduced distance between neighboring spin objects. This in turn induces stronger repulsive forces~\cite{Zhang2015,Castro2020}, which give rise to the observed higher breathing mode frequency. Furthermore, not only the amplitude and frequency of the breathing mode are increased, but also the equilibrium position of oscillation is shifted. In Fig.~\ref{fig:Data}a the minima, at which the spin objects reach their maximum size, almost overlap, while the maxima, indicating the minimum size of the spin objects, are clearly separated. In other words, the maximum size of the spin objects during the breathing mode oscillation is limited, which we attribute to the collective expansion and contraction of the individual spin objects within the dense, ordered lattice structure. In fact, the closely packed lattice prohibits further expansion of the spin objects as they feel their corresponding neighbors, an effect most likely not relevant in the simulation due to the less dense skyrmion packing.    

\section{Conclusion}
In summary, we found that applying an in-plane magnetic set field prior to establishing a specific out-of-plane magnetic field in Fe/Gd multilayer films modifies both the amplitude and frequency of laser-induced breathing modes. Using LTEM, MFM, and micromagnetic simulations, we explained the observed changes by a persistent modification of ground state magnetic spin textures by the in-plane magnetic set field, allowing for the creation of a larger amount of spin objects within the BSK lattice structure. In more detail, our studies indicate the formation of a more dense, hexagonally ordered single skyrmion lattice. The larger breathing amplitude then follows from the enhanced density of spin objects, since more objects contribute to the collective breathing mode. Similarly, the experimentally observed higher frequency can be related to increased repulsive forces between neighboring objects due to their reduced distance. The shift in the equilibrium position of the breathing may be indicative of the relevance of interactions between the different spin objects, limiting their maximum expansion within the lattice. In total, our observations show a pathway for actively controlling and harnessing the density and ordering of dipolar-stabilized spin objects in multilayer thin films, potentially allowing even for topology control, which might be of interest for future applications. 

\subsection*{Experimental Section}
\subsubsection*{Time-Resolved Magneto-Optical Kerr Effect (TR-MOKE)}
Magnetization dynamics $\Delta M(t)$ of a [Fe($0.35$\,nm)/Gd($0.40$\,nm)]$_{160}$ thin film, deposited onto a thermally-oxidized Si(100) substrate, were measured using a bichromatic pump-probe setup. Sub-ps laser pulses at a repetition rate of $50$\,kHz were provided by a fiber amplifier system. These pulses were compressed to less than $40$\,fs using spectral broadening in a gas-filled hollow core fiber and compression by chirped mirrors. The fundamental light pulses with central wavelength of $1030$\,nm were used as pump pulses, while their second harmonic at $515$\,nm represent the probe pulses. External out-of-plane magnetic fields up to $\mu_0H_{\mathrm{oop}}=0.5$\,T could be applied in-between the poles of a variable-gap electromagnet. There, we added another electromagnet with fixed gap to provide an external in-plane magnetic field of up to $\mu_0H_{\mathrm{ip}}=0.2$\,T at the centered sample position. A balanced bridge detector was used to measure laser-induced polarization changes, which can be related to magnetization changes. The data was acquired by a $250$\,MHz digitzer card in a chopped measurement mode, allowing for maximum scope of data post processing. Note that the setup makes use of a stroboscopic pump-probe method, i.e., only reversible dynamics are accessible and one data point for a given time-delay and set out-of-plane magnetic field represents the average signal over thousands of pump-probe cycles. 

\subsubsection*{Micromagnetic Simulations}
We used \texttt{magnum.np}~\cite{Bruckner2023} to perform micromagnetic simulations using the finite-difference method. The simulations were carried out on the [Fe($0.35$\,nm)/Gd($0.40$\,nm)]$_{160}$ sample using a box of dimensions ($512\times 512\times 10$) where each cell had the size ($10\times10\times12{\rm\,nm^3}$). Two-dimensional periodic boundary conditions were used with 10 repetitions in each directions. The material parameters at $T=\SI{300}{K}$ are: $M_s = 340\mathrm{kA/m}$, $K_u = 40\,\mathrm{kJ/m^3}$ and $M_s = 6\,\mathrm{pJ/m}$, which have yielded excellent agreement with experiments in the past~\cite{Titze2024b, Titze2025}. 

\paragraph{Static Simulations.} We start from a random magnetization state, selected by a manual seed, and proceed with the following simulation protocol. First, we relax the magnetization at vanishing magnetic fields by numerically integrating the Landau-Lifshitz-Gilbert (LLG) equation for $5\,\rm{ns}$ by considering the contributions from demagnetization, exchange and anisotropy energies. Thermal fluctuations are neglected at this state. Then, an in-plane magnetic field is applied ($\mu_0H_{\rm ip} = 25$--$100\,\rm{mT}$), and the magnetic configuration is relaxed with the additional Zeeman energy. In a third step, the magnetization is relaxed at vanishing fields. From here on, we include a time-dependent external out-of-plane magnetic field into the energy terms, where the $\mu_0H_z$ component of the field increases from $0\,\rm{mT}$ to $300\,\rm{mT}$ in $6$\,{\textmu}s. The magnetization state is then recorded every $10\,\rm{ns}$.
\paragraph{Dynamic Simulations.} We follow closely the approach adapted in Ref.~\cite{Titze2024b}, where an effective temperature is calculated from the time-dependent change in total magnetization, which was recorded experimentally. We choose $\mu_0H_z = 220~\rm{mT}$, and $\alpha = 0.02$. The initial magnetic state is obtained from the static simulations described above. Thermal fluctuations are now considered~\cite{Bruckner2023}.  First, the magnetization is relaxed under the consideration of the thermal fluctuations at $T=300~\rm{K}$. Then the laser-excitation is mimicked by following the experimental demagnetization curve. The magnetization is recorded for $3~\rm{ns}$, until an equilibrium state is reached. The excitation scheme is repeated 8 times, to investigate the influence of the stroboscopic measurement technique employed in the experiments.

\subsubsection*{Lorentz Transmission Electron Microscopy}
The LTEM imaging of the magnetic spin structure in the thin films deposited on the membranes was realized using a JEOL NEOARM-200F TEM system in Fresnel mode. A tilting holder was used in order to change the angle between the sample surface and the applied magnetic field. We employed the following experimental procedure. First, the disordered stripe state was set in the sample by an out-of-plane demagnetization routine. We then imaged the magnetic states at increasing applied out-of-plane fields. After the sample reached the saturated state, the experiment was reset by returning to zero-field. Now, the aligned stripe state was set by tilting the sample by $30^{\circ}$ and applying a moderate field, which now entails a considerable in-plane component. After returning to zero-field and zero-tilt, we then imaged the following evolution of the magnetic state by increasing the applied magnetic field. With this protocol, the evolution of both states could be imaged for the same membrane area, which was chosen due to low membrane buckling effects and mostly planar arrangement.

\subsubsection*{Magnetic Force Microscopy}
A Bruker Dimension Icon System was used to perform MFM Imaging employing low-moment tips. The applied magnetic field was realized by placing a permanent magnet (N45) and spacers beneath the sample during scanning. Prior to the scans with an applied field, appropriate amounts of spacers were placed between the permanent magnet and the sample and the field strengths at the position of the scan were measured with a Hall sensor. This way, two setups were found for 160\,mT and 220\,mT, corresponding to the mixed state of stripe domains and cylindrical spin objects, and a BSK lattice, respectively. In order to set the disordered stripe state in the sample, analogously to the LTEM experiment, an out-of-plane demagnetization routine was applied to the [Fe($0.35$\,nm)/Gd($0.40$\,nm)]$_{160}$-film on a rigid silicon substrate. The aligned stripe state was set by applying an in-plane field to the sample with a permanent magnet in its proximity. Prior to every scan the desired magnetic state was set and the sample was placed onto the magnet and spaceholders to apply the field.

\begin{acknowledgments}
T.T.\,and D.S.\,gratefully acknowledge funding by the Deutsche Forschungsgemeinschaft (DFG, German Research Foundation), Grant No.\,217133147 (SFB1073, project A02). T.S. and M.A. gratefully acknowledge funding by the Deutsche Forschungsgemeinschaft (DFG, German Research Foundation) Grant No. 507821284 and 540566574. S.K.\,and D.Su.\,acknowledge the Austrian Science Fund (FWF) for support through Grant No.\,I\,6267 (CHIRALSPIN). S.K.\,, and A.F.P.\,acknowledge funding from the European Research Council
(ERC) under the European Union’s Horizon 2020 research
and innovation programme, Grant Agreement No. 101001290
(3DNANOMAG). We acknowledge Vienna Scientific Cluster
(VSC) for awarding this project access to the LEONARDO
supercomputer, owned by the EuroHPC Joint Undertaking,
hosted by CINECA (Italy) and the LEONARDO consortium.
\end{acknowledgments}

\subsection*{Author Contributions}
S. Mathias and D. Steil conceived the project. T. Titze and M. Matthies performed
the time-resolved Kerr effect studies and analyzed the time-resolved magnetization data
under the supervision of S. Mathias and D. Steil. Sample growth, MFM measurements, and LTEM measurements were performed by T. Schmidt under the supervision of M. Albrecht. S. Koraltan performed the micromagnetic simulations under the supervision of D. Suess and A. Fern\'{a}ndez-Pacheco. The simulations were evaluated by S. Koraltan and T. Titze with support from D. Steil. T. Titze and D. Steil wrote the manuscript with input from all authors.

\subsection*{Data availability statement}
The data that support the findings of this study are available from the authors upon reasonable request.

\subsection*{Appendix}
\subsubsection*{Membrane Buckling Effects}
\begin{figure}[t]
     \centering
     \includegraphics[width=1\columnwidth]{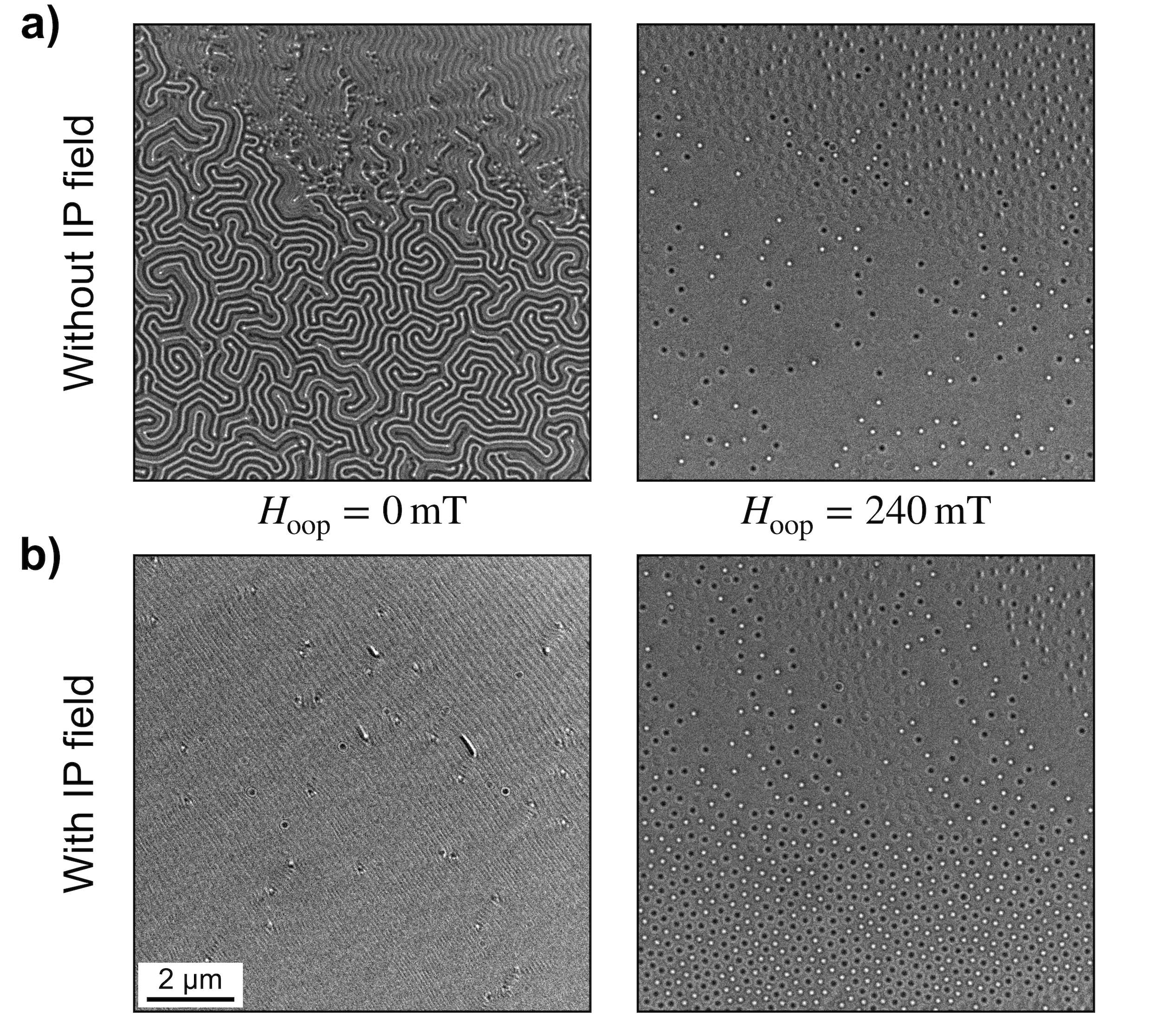}
     \caption{\textbf{Magnetic spin textures in the presence of membrane buckling.} \textbf{a)} Magnetic spin textures obtained from LTEM using two different out-of-plane magnetic fields. \textbf{b)} Same as in a), but an in-plane magnetic field was applied before increasing the out-of-plane magnetic field. Membrane buckling effects are prominent in the top right corner, giving rise to the formation of topologically trivial bubbles at high out-of-plane magnetic fields. Note that the images show a similar field of view, however, minor distortions are expected due to the high magnetic field.}
     \label{fig:Membrane}
\end{figure}
The [Fe($0.35$\,nm)/Gd($0.40$\,nm)]$_{120}$ sample used for LTEM imaging was prepared on a $30$\,nm thick Si$_3$N$_4$, which introduces strain and thus intrinsic membrane wrinkling, which induces a tilted
anisotropy axis of the magnetic film and effective in-plane field components, when an out-of-plane field is applied. As a result, the morphology of the magnetic spin textures is drastically changed. In regions with strong membrane buckling, predominantly non-chiral domain walls, respectively topologically trivial magnetic bubbles appear, as shown in the top right corner of Fig.~\ref{fig:Membrane}a. Similar effects are observed in Fig.~\ref{fig:Membrane}b in this region after the sample has been exposed to an in-plane magnetic field, meaning that the membrane buckling effects dominate in this region.

\subsubsection*{Metastable Bubble Lattice}
\begin{figure}[t]
     \centering
     \includegraphics[width=1\columnwidth]{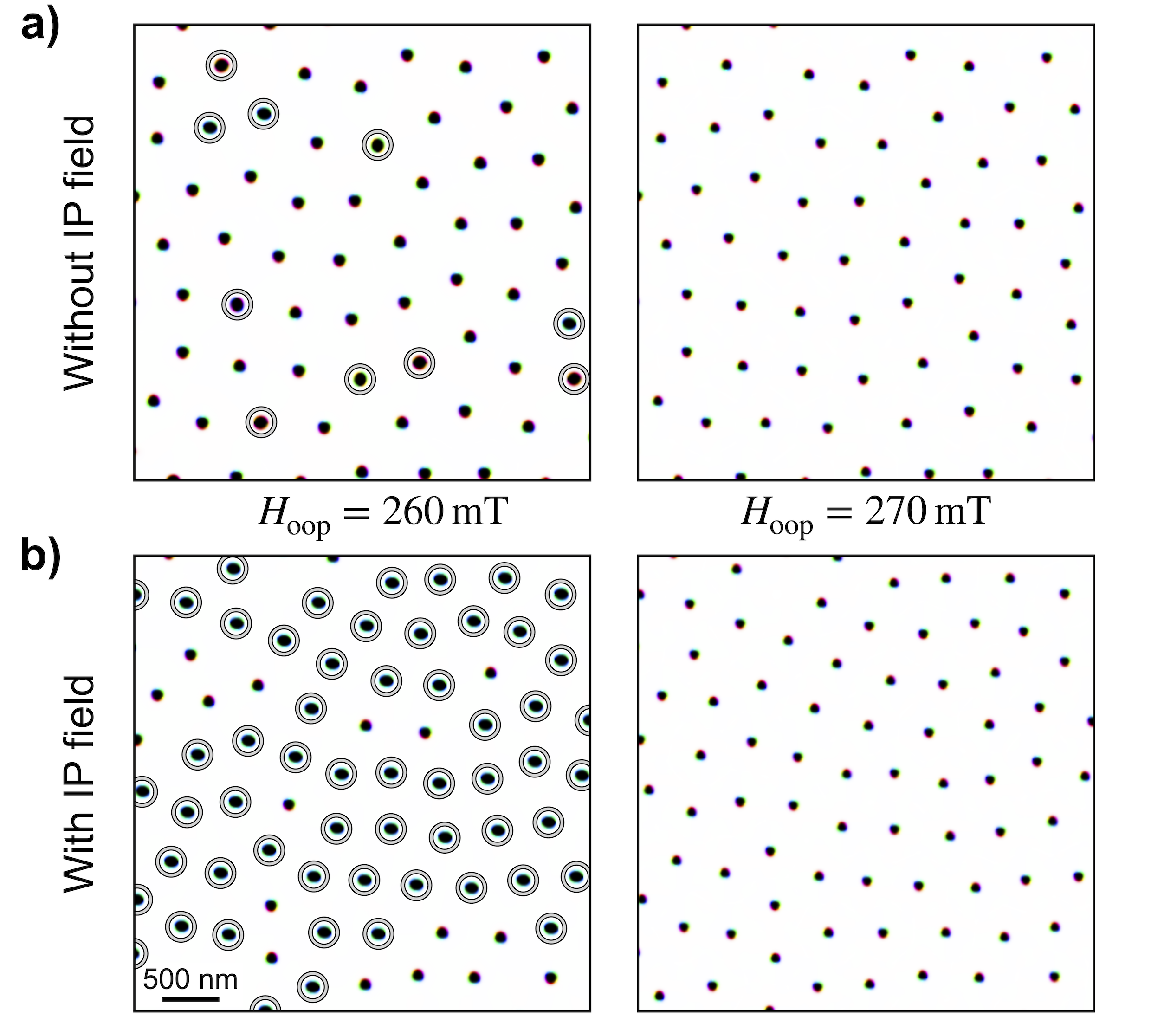}
     \caption{\textbf{Formation of a single skyrmion lattice at high out-of-plane magnetic fields.} \textbf{a)} Magnetic spin textures obtained from micromagnetic simulations using two different out-of-plane magnetic fields. \textbf{b)} Same as in a), but an in-plane magnetic set field was applied before increasing the out-of-plane magnetic field. Topologically trivial magnetic bubbles are highlighted by gray circles for better visibility.}
     \label{fig:SimSingleSkyrmion}
\end{figure}
Examining the evolution of magnetic spin textures in Fig.~\ref{fig:Simulation} under an applied in-plane magnetic set field, we observe that the initially aligned stripe domains predominantly transform into topologically trivial bubbles as the out-of-plane field is increased. Figure~\ref{fig:SimSingleSkyrmion} displays the magnetic spin textures at two elevated magnetic fields for the two cases with and without prior in-plane magnetic set field. Interestingly, all bubbles transform into topologically non-trivial skyrmions when the magnetic field is slightly increased, highlighting the metastable character of the trivial bubbles. Similar transformations can also be induced at lower magnetic fields through weak laser excitation, which effectively mimics small temperature variations and thereby modifies the magnetic properties. The transformation itself is mediated by the formation of Bloch points, as elaborated in the main text.
\newpage
\bibliographystyle{apsrev4-2}
\bibliography{references}

\begin{thebibliography}{37}%
\makeatletter
\providecommand \@ifxundefined [1]{%
 \@ifx{#1\undefined}
}%
\providecommand \@ifnum [1]{%
 \ifnum #1\expandafter \@firstoftwo
 \else \expandafter \@secondoftwo
 \fi
}%
\providecommand \@ifx [1]{%
 \ifx #1\expandafter \@firstoftwo
 \else \expandafter \@secondoftwo
 \fi
}%
\providecommand \natexlab [1]{#1}%
\providecommand \enquote  [1]{``#1''}%
\providecommand \bibnamefont  [1]{#1}%
\providecommand \bibfnamefont [1]{#1}%
\providecommand \citenamefont [1]{#1}%
\providecommand \href@noop [0]{\@secondoftwo}%
\providecommand \href [0]{\begingroup \@sanitize@url \@href}%
\providecommand \@href[1]{\@@startlink{#1}\@@href}%
\providecommand \@@href[1]{\endgroup#1\@@endlink}%
\providecommand \@sanitize@url [0]{\catcode `\\12\catcode `\$12\catcode
  `\&12\catcode `\#12\catcode `\^12\catcode `\_12\catcode `\%12\relax}%
\providecommand \@@startlink[1]{}%
\providecommand \@@endlink[0]{}%
\providecommand \url  [0]{\begingroup\@sanitize@url \@url }%
\providecommand \@url [1]{\endgroup\@href {#1}{\urlprefix }}%
\providecommand \urlprefix  [0]{URL }%
\providecommand \Eprint [0]{\href }%
\providecommand \doibase [0]{https://doi.org/}%
\providecommand \selectlanguage [0]{\@gobble}%
\providecommand \bibinfo  [0]{\@secondoftwo}%
\providecommand \bibfield  [0]{\@secondoftwo}%
\providecommand \translation [1]{[#1]}%
\providecommand \BibitemOpen [0]{}%
\providecommand \bibitemStop [0]{}%
\providecommand \bibitemNoStop [0]{.\EOS\space}%
\providecommand \EOS [0]{\spacefactor3000\relax}%
\providecommand \BibitemShut  [1]{\csname bibitem#1\endcsname}%
\let\auto@bib@innerbib\@empty
\bibitem [{\citenamefont {Mühlbauer}\ \emph {et~al.}(2009)\citenamefont
  {Mühlbauer}, \citenamefont {Binz}, \citenamefont {Jonietz}, \citenamefont
  {Pfleiderer}, \citenamefont {Rosch}, \citenamefont {Neubauer}, \citenamefont
  {Georgii},\ and\ \citenamefont {Böni}}]{Muehlbauer2009}%
  \BibitemOpen
  \bibfield  {author} {\bibinfo {author} {\bibfnamefont {S.}~\bibnamefont
  {Mühlbauer}}, \bibinfo {author} {\bibfnamefont {B.}~\bibnamefont {Binz}},
  \bibinfo {author} {\bibfnamefont {F.}~\bibnamefont {Jonietz}}, \bibinfo
  {author} {\bibfnamefont {C.}~\bibnamefont {Pfleiderer}}, \bibinfo {author}
  {\bibfnamefont {A.}~\bibnamefont {Rosch}}, \bibinfo {author} {\bibfnamefont
  {A.}~\bibnamefont {Neubauer}}, \bibinfo {author} {\bibfnamefont
  {R.}~\bibnamefont {Georgii}},\ and\ \bibinfo {author} {\bibfnamefont
  {P.}~\bibnamefont {Böni}},\ }\href {https://doi.org/10.1126/science.1166767}
  {\bibfield  {journal} {\bibinfo  {journal} {Science}\ }\textbf {\bibinfo
  {volume} {323}},\ \bibinfo {pages} {915} (\bibinfo {year}
  {2009})}\BibitemShut {NoStop}%
\bibitem [{\citenamefont {Nagaosa}\ and\ \citenamefont
  {Tokura}(2013)}]{Nagaosa2013}%
  \BibitemOpen
  \bibfield  {author} {\bibinfo {author} {\bibfnamefont {N.}~\bibnamefont
  {Nagaosa}}\ and\ \bibinfo {author} {\bibfnamefont {Y.}~\bibnamefont
  {Tokura}},\ }\href {https://doi.org/10.1038/nnano.2013.243} {\bibfield
  {journal} {\bibinfo  {journal} {Nature Nanotechnology}\ }\textbf {\bibinfo
  {volume} {8}},\ \bibinfo {pages} {899} (\bibinfo {year} {2013})}\BibitemShut
  {NoStop}%
\bibitem [{\citenamefont {Finocchio}\ \emph {et~al.}(2016)\citenamefont
  {Finocchio}, \citenamefont {B{\"u}ttner}, \citenamefont {Tomasello},
  \citenamefont {Carpentieri},\ and\ \citenamefont
  {Kl{\"a}ui}}]{Finocchio2016}%
  \BibitemOpen
  \bibfield  {author} {\bibinfo {author} {\bibfnamefont {G.}~\bibnamefont
  {Finocchio}}, \bibinfo {author} {\bibfnamefont {F.}~\bibnamefont
  {B{\"u}ttner}}, \bibinfo {author} {\bibfnamefont {R.}~\bibnamefont
  {Tomasello}}, \bibinfo {author} {\bibfnamefont {M.}~\bibnamefont
  {Carpentieri}},\ and\ \bibinfo {author} {\bibfnamefont {M.}~\bibnamefont
  {Kl{\"a}ui}},\ }\href@noop {} {\bibfield  {journal} {\bibinfo  {journal}
  {Journal of Physics D: Applied Physics}\ }\textbf {\bibinfo {volume} {49}},\
  \bibinfo {pages} {423001} (\bibinfo {year} {2016})}\BibitemShut {NoStop}%
\bibitem [{\citenamefont {Fert}\ \emph {et~al.}(2017)\citenamefont {Fert},
  \citenamefont {Reyren},\ and\ \citenamefont {Cros}}]{Fert2017}%
  \BibitemOpen
  \bibfield  {author} {\bibinfo {author} {\bibfnamefont {A.}~\bibnamefont
  {Fert}}, \bibinfo {author} {\bibfnamefont {N.}~\bibnamefont {Reyren}},\ and\
  \bibinfo {author} {\bibfnamefont {V.}~\bibnamefont {Cros}},\ }\href@noop {}
  {\bibfield  {journal} {\bibinfo  {journal} {Nature Reviews Materials}\
  }\textbf {\bibinfo {volume} {2}},\ \bibinfo {pages} {1} (\bibinfo {year}
  {2017})}\BibitemShut {NoStop}%
\bibitem [{\citenamefont {Grollier}\ \emph {et~al.}(2020)\citenamefont
  {Grollier}, \citenamefont {Querlioz}, \citenamefont {Camsari}, \citenamefont
  {Everschor-Sitte}, \citenamefont {Fukami},\ and\ \citenamefont
  {Stiles}}]{Grollier2020}%
  \BibitemOpen
  \bibfield  {author} {\bibinfo {author} {\bibfnamefont {J.}~\bibnamefont
  {Grollier}}, \bibinfo {author} {\bibfnamefont {D.}~\bibnamefont {Querlioz}},
  \bibinfo {author} {\bibfnamefont {K.~Y.}\ \bibnamefont {Camsari}}, \bibinfo
  {author} {\bibfnamefont {K.}~\bibnamefont {Everschor-Sitte}}, \bibinfo
  {author} {\bibfnamefont {S.}~\bibnamefont {Fukami}},\ and\ \bibinfo {author}
  {\bibfnamefont {M.~D.}\ \bibnamefont {Stiles}},\ }\href
  {https://doi.org/10.1038/s41928-019-0360-9} {\bibfield  {journal} {\bibinfo
  {journal} {Nature Electronics}\ }\textbf {\bibinfo {volume} {3}},\ \bibinfo
  {pages} {360} (\bibinfo {year} {2020})}\BibitemShut {NoStop}%
\bibitem [{\citenamefont {Yu}\ \emph {et~al.}(2021)\citenamefont {Yu},
  \citenamefont {Xiao},\ and\ \citenamefont {Schultheiss}}]{Yu2021}%
  \BibitemOpen
  \bibfield  {author} {\bibinfo {author} {\bibfnamefont {H.}~\bibnamefont
  {Yu}}, \bibinfo {author} {\bibfnamefont {J.}~\bibnamefont {Xiao}},\ and\
  \bibinfo {author} {\bibfnamefont {H.}~\bibnamefont {Schultheiss}},\ }\href
  {https://doi.org/https://doi.org/10.1016/j.physrep.2020.12.004} {\bibfield
  {journal} {\bibinfo  {journal} {Physics Reports}\ }\textbf {\bibinfo {volume}
  {905}},\ \bibinfo {pages} {1} (\bibinfo {year} {2021})},\ \bibinfo {note}
  {magnetic texture based magnonics}\BibitemShut {NoStop}%
\bibitem [{\citenamefont {Heinze}\ \emph {et~al.}(2011)\citenamefont {Heinze},
  \citenamefont {von Bergmann}, \citenamefont {Menzel}, \citenamefont {Brede},
  \citenamefont {Kubetzka}, \citenamefont {Wiesendanger}, \citenamefont
  {Bihlmayer},\ and\ \citenamefont {Bl{\"u}gel}}]{Heinze2011}%
  \BibitemOpen
  \bibfield  {author} {\bibinfo {author} {\bibfnamefont {S.}~\bibnamefont
  {Heinze}}, \bibinfo {author} {\bibfnamefont {K.}~\bibnamefont {von
  Bergmann}}, \bibinfo {author} {\bibfnamefont {M.}~\bibnamefont {Menzel}},
  \bibinfo {author} {\bibfnamefont {J.}~\bibnamefont {Brede}}, \bibinfo
  {author} {\bibfnamefont {A.}~\bibnamefont {Kubetzka}}, \bibinfo {author}
  {\bibfnamefont {R.}~\bibnamefont {Wiesendanger}}, \bibinfo {author}
  {\bibfnamefont {G.}~\bibnamefont {Bihlmayer}},\ and\ \bibinfo {author}
  {\bibfnamefont {S.}~\bibnamefont {Bl{\"u}gel}},\ }\href@noop {} {\bibfield
  {journal} {\bibinfo  {journal} {Nat. Phys.}\ }\textbf {\bibinfo {volume}
  {7}},\ \bibinfo {pages} {713} (\bibinfo {year} {2011})}\BibitemShut {NoStop}%
\bibitem [{\citenamefont {Chen}\ \emph {et~al.}(2015)\citenamefont {Chen},
  \citenamefont {Mascaraque}, \citenamefont {N'Diaye},\ and\ \citenamefont
  {Schmid}}]{Chen2015}%
  \BibitemOpen
  \bibfield  {author} {\bibinfo {author} {\bibfnamefont {G.}~\bibnamefont
  {Chen}}, \bibinfo {author} {\bibfnamefont {A.}~\bibnamefont {Mascaraque}},
  \bibinfo {author} {\bibfnamefont {A.~T.}\ \bibnamefont {N'Diaye}},\ and\
  \bibinfo {author} {\bibfnamefont {A.~K.}\ \bibnamefont {Schmid}},\ }\href
  {https://doi.org/10.1063/1.4922726} {\bibfield  {journal} {\bibinfo
  {journal} {Appl. Phys. Lett.}\ }\textbf {\bibinfo {volume} {106}},\ \bibinfo
  {pages} {242404} (\bibinfo {year} {2015})}\BibitemShut {NoStop}%
\bibitem [{\citenamefont {Woo}\ \emph {et~al.}(2016)\citenamefont {Woo},
  \citenamefont {Litzius}, \citenamefont {Kr{\"u}ger}, \citenamefont {Im},
  \citenamefont {Caretta}, \citenamefont {Richter}, \citenamefont {Mann},
  \citenamefont {Krone}, \citenamefont {Reeve}, \citenamefont {Weigand},
  \citenamefont {Agrawal}, \citenamefont {Lemesh}, \citenamefont {Mawass},
  \citenamefont {Fischer}, \citenamefont {Kl{\"a}ui},\ and\ \citenamefont
  {Beach}}]{Woo2016}%
  \BibitemOpen
  \bibfield  {author} {\bibinfo {author} {\bibfnamefont {S.}~\bibnamefont
  {Woo}}, \bibinfo {author} {\bibfnamefont {K.}~\bibnamefont {Litzius}},
  \bibinfo {author} {\bibfnamefont {B.}~\bibnamefont {Kr{\"u}ger}}, \bibinfo
  {author} {\bibfnamefont {M.-Y.}\ \bibnamefont {Im}}, \bibinfo {author}
  {\bibfnamefont {L.}~\bibnamefont {Caretta}}, \bibinfo {author} {\bibfnamefont
  {K.}~\bibnamefont {Richter}}, \bibinfo {author} {\bibfnamefont
  {M.}~\bibnamefont {Mann}}, \bibinfo {author} {\bibfnamefont {A.}~\bibnamefont
  {Krone}}, \bibinfo {author} {\bibfnamefont {R.~M.}\ \bibnamefont {Reeve}},
  \bibinfo {author} {\bibfnamefont {M.}~\bibnamefont {Weigand}}, \bibinfo
  {author} {\bibfnamefont {P.}~\bibnamefont {Agrawal}}, \bibinfo {author}
  {\bibfnamefont {I.}~\bibnamefont {Lemesh}}, \bibinfo {author} {\bibfnamefont
  {M.-A.}\ \bibnamefont {Mawass}}, \bibinfo {author} {\bibfnamefont
  {P.}~\bibnamefont {Fischer}}, \bibinfo {author} {\bibfnamefont
  {M.}~\bibnamefont {Kl{\"a}ui}},\ and\ \bibinfo {author} {\bibfnamefont
  {G.~S.~D.}\ \bibnamefont {Beach}},\ }\href {https://doi.org/10.1038/nmat4593}
  {\bibfield  {journal} {\bibinfo  {journal} {Nat. Mater.}\ }\textbf {\bibinfo
  {volume} {15}},\ \bibinfo {pages} {501} (\bibinfo {year} {2016})}\BibitemShut
  {NoStop}%
\bibitem [{\citenamefont {M\"unzer}\ \emph {et~al.}(2010)\citenamefont
  {M\"unzer}, \citenamefont {Neubauer}, \citenamefont {Adams}, \citenamefont
  {M\"uhlbauer}, \citenamefont {Franz}, \citenamefont {Jonietz}, \citenamefont
  {Georgii}, \citenamefont {B\"oni}, \citenamefont {Pedersen}, \citenamefont
  {Schmidt}, \citenamefont {Rosch},\ and\ \citenamefont
  {Pfleiderer}}]{Muenzer2010}%
  \BibitemOpen
  \bibfield  {author} {\bibinfo {author} {\bibfnamefont {W.}~\bibnamefont
  {M\"unzer}}, \bibinfo {author} {\bibfnamefont {A.}~\bibnamefont {Neubauer}},
  \bibinfo {author} {\bibfnamefont {T.}~\bibnamefont {Adams}}, \bibinfo
  {author} {\bibfnamefont {S.}~\bibnamefont {M\"uhlbauer}}, \bibinfo {author}
  {\bibfnamefont {C.}~\bibnamefont {Franz}}, \bibinfo {author} {\bibfnamefont
  {F.}~\bibnamefont {Jonietz}}, \bibinfo {author} {\bibfnamefont
  {R.}~\bibnamefont {Georgii}}, \bibinfo {author} {\bibfnamefont
  {P.}~\bibnamefont {B\"oni}}, \bibinfo {author} {\bibfnamefont
  {B.}~\bibnamefont {Pedersen}}, \bibinfo {author} {\bibfnamefont
  {M.}~\bibnamefont {Schmidt}}, \bibinfo {author} {\bibfnamefont
  {A.}~\bibnamefont {Rosch}},\ and\ \bibinfo {author} {\bibfnamefont
  {C.}~\bibnamefont {Pfleiderer}},\ }\href
  {https://doi.org/10.1103/PhysRevB.81.041203} {\bibfield  {journal} {\bibinfo
  {journal} {Phys. Rev. B}\ }\textbf {\bibinfo {volume} {81}},\ \bibinfo
  {pages} {041203} (\bibinfo {year} {2010})}\BibitemShut {NoStop}%
\bibitem [{\citenamefont {Yu}\ \emph {et~al.}(2010)\citenamefont {Yu},
  \citenamefont {Onose}, \citenamefont {Kanazawa}, \citenamefont {Park},
  \citenamefont {Han}, \citenamefont {Matsui}, \citenamefont {Nagaosa},\ and\
  \citenamefont {Tokura}}]{Yu2010}%
  \BibitemOpen
  \bibfield  {author} {\bibinfo {author} {\bibfnamefont {X.~Z.}\ \bibnamefont
  {Yu}}, \bibinfo {author} {\bibfnamefont {Y.}~\bibnamefont {Onose}}, \bibinfo
  {author} {\bibfnamefont {N.}~\bibnamefont {Kanazawa}}, \bibinfo {author}
  {\bibfnamefont {J.~H.}\ \bibnamefont {Park}}, \bibinfo {author}
  {\bibfnamefont {J.~H.}\ \bibnamefont {Han}}, \bibinfo {author} {\bibfnamefont
  {Y.}~\bibnamefont {Matsui}}, \bibinfo {author} {\bibfnamefont
  {N.}~\bibnamefont {Nagaosa}},\ and\ \bibinfo {author} {\bibfnamefont
  {Y.}~\bibnamefont {Tokura}},\ }\href {https://doi.org/10.1038/nature09124}
  {\bibfield  {journal} {\bibinfo  {journal} {Nature}\ }\textbf {\bibinfo
  {volume} {465}},\ \bibinfo {pages} {901} (\bibinfo {year}
  {2010})}\BibitemShut {NoStop}%
\bibitem [{\citenamefont {Yu}\ \emph {et~al.}(2012)\citenamefont {Yu},
  \citenamefont {Mostovoy}, \citenamefont {Tokunaga}, \citenamefont {Zhang},
  \citenamefont {Kimoto}, \citenamefont {Matsui}, \citenamefont {Kaneko},
  \citenamefont {Nagaosa},\ and\ \citenamefont {Tokura}}]{Yu2012}%
  \BibitemOpen
  \bibfield  {author} {\bibinfo {author} {\bibfnamefont {X.}~\bibnamefont
  {Yu}}, \bibinfo {author} {\bibfnamefont {M.}~\bibnamefont {Mostovoy}},
  \bibinfo {author} {\bibfnamefont {Y.}~\bibnamefont {Tokunaga}}, \bibinfo
  {author} {\bibfnamefont {W.}~\bibnamefont {Zhang}}, \bibinfo {author}
  {\bibfnamefont {K.}~\bibnamefont {Kimoto}}, \bibinfo {author} {\bibfnamefont
  {Y.}~\bibnamefont {Matsui}}, \bibinfo {author} {\bibfnamefont
  {Y.}~\bibnamefont {Kaneko}}, \bibinfo {author} {\bibfnamefont
  {N.}~\bibnamefont {Nagaosa}},\ and\ \bibinfo {author} {\bibfnamefont
  {Y.}~\bibnamefont {Tokura}},\ }\href
  {https://doi.org/10.1073/pnas.1118496109} {\bibfield  {journal} {\bibinfo
  {journal} {Proc. Natl. Acad. Sci.}\ }\textbf {\bibinfo {volume} {109}},\
  \bibinfo {pages} {8856} (\bibinfo {year} {2012})}\BibitemShut {NoStop}%
\bibitem [{\citenamefont {Lee}\ \emph {et~al.}(2016)\citenamefont {Lee},
  \citenamefont {Chess}, \citenamefont {Montoya}, \citenamefont {Shi},
  \citenamefont {Tamura}, \citenamefont {Mishra}, \citenamefont {Fischer},
  \citenamefont {McMorran}, \citenamefont {Sinha}, \citenamefont {Fullerton},
  \citenamefont {Kevan},\ and\ \citenamefont {Roy}}]{Lee2016FeGd}%
  \BibitemOpen
  \bibfield  {author} {\bibinfo {author} {\bibfnamefont {J.~C.~T.}\
  \bibnamefont {Lee}}, \bibinfo {author} {\bibfnamefont {J.~J.}\ \bibnamefont
  {Chess}}, \bibinfo {author} {\bibfnamefont {S.~A.}\ \bibnamefont {Montoya}},
  \bibinfo {author} {\bibfnamefont {X.}~\bibnamefont {Shi}}, \bibinfo {author}
  {\bibfnamefont {N.}~\bibnamefont {Tamura}}, \bibinfo {author} {\bibfnamefont
  {S.~K.}\ \bibnamefont {Mishra}}, \bibinfo {author} {\bibfnamefont
  {P.}~\bibnamefont {Fischer}}, \bibinfo {author} {\bibfnamefont {B.~J.}\
  \bibnamefont {McMorran}}, \bibinfo {author} {\bibfnamefont {S.~K.}\
  \bibnamefont {Sinha}}, \bibinfo {author} {\bibfnamefont {E.~E.}\ \bibnamefont
  {Fullerton}}, \bibinfo {author} {\bibfnamefont {S.~D.}\ \bibnamefont
  {Kevan}},\ and\ \bibinfo {author} {\bibfnamefont {S.}~\bibnamefont {Roy}},\
  }\href {https://doi.org/10.1063/1.4955462} {\bibfield  {journal} {\bibinfo
  {journal} {Appl. Phys. Lett.}\ }\textbf {\bibinfo {volume} {109}},\ \bibinfo
  {pages} {022402} (\bibinfo {year} {2016})}\BibitemShut {NoStop}%
\bibitem [{\citenamefont {Desautels}\ \emph {et~al.}(2019)\citenamefont
  {Desautels}, \citenamefont {DeBeer-Schmitt}, \citenamefont {Montoya},
  \citenamefont {Borchers}, \citenamefont {Je}, \citenamefont {Tang},
  \citenamefont {Im}, \citenamefont {Fitzsimmons}, \citenamefont {Fullerton},\
  and\ \citenamefont {Gilbert}}]{Desautels2019}%
  \BibitemOpen
  \bibfield  {author} {\bibinfo {author} {\bibfnamefont {R.~D.}\ \bibnamefont
  {Desautels}}, \bibinfo {author} {\bibfnamefont {L.}~\bibnamefont
  {DeBeer-Schmitt}}, \bibinfo {author} {\bibfnamefont {S.~A.}\ \bibnamefont
  {Montoya}}, \bibinfo {author} {\bibfnamefont {J.~A.}\ \bibnamefont
  {Borchers}}, \bibinfo {author} {\bibfnamefont {S.-G.}\ \bibnamefont {Je}},
  \bibinfo {author} {\bibfnamefont {N.}~\bibnamefont {Tang}}, \bibinfo {author}
  {\bibfnamefont {M.-Y.}\ \bibnamefont {Im}}, \bibinfo {author} {\bibfnamefont
  {M.~R.}\ \bibnamefont {Fitzsimmons}}, \bibinfo {author} {\bibfnamefont
  {E.~E.}\ \bibnamefont {Fullerton}},\ and\ \bibinfo {author} {\bibfnamefont
  {D.~A.}\ \bibnamefont {Gilbert}},\ }\href
  {https://doi.org/10.1103/PhysRevMaterials.3.104406} {\bibfield  {journal}
  {\bibinfo  {journal} {Phys. Rev. Mater.}\ }\textbf {\bibinfo {volume} {3}},\
  \bibinfo {pages} {104406} (\bibinfo {year} {2019})}\BibitemShut {NoStop}%
\bibitem [{\citenamefont {Zhang}\ \emph {et~al.}(2020)\citenamefont {Zhang},
  \citenamefont {Zhang}, \citenamefont {Chen}, \citenamefont {Guang},
  \citenamefont {Zeng}, \citenamefont {Yu}, \citenamefont {Zhang},
  \citenamefont {Liu}, \citenamefont {Feng}, \citenamefont {Zhao},
  \citenamefont {Zhou}, \citenamefont {Qiu}, \citenamefont {Han}, \citenamefont
  {Peng},\ and\ \citenamefont {Zhang}}]{Zhang2020c}%
  \BibitemOpen
  \bibfield  {author} {\bibinfo {author} {\bibfnamefont {J.}~\bibnamefont
  {Zhang}}, \bibinfo {author} {\bibfnamefont {X.}~\bibnamefont {Zhang}},
  \bibinfo {author} {\bibfnamefont {H.}~\bibnamefont {Chen}}, \bibinfo {author}
  {\bibfnamefont {Y.}~\bibnamefont {Guang}}, \bibinfo {author} {\bibfnamefont
  {X.}~\bibnamefont {Zeng}}, \bibinfo {author} {\bibfnamefont {G.}~\bibnamefont
  {Yu}}, \bibinfo {author} {\bibfnamefont {S.}~\bibnamefont {Zhang}}, \bibinfo
  {author} {\bibfnamefont {Y.}~\bibnamefont {Liu}}, \bibinfo {author}
  {\bibfnamefont {J.}~\bibnamefont {Feng}}, \bibinfo {author} {\bibfnamefont
  {Y.}~\bibnamefont {Zhao}}, \bibinfo {author} {\bibfnamefont {Y.}~\bibnamefont
  {Zhou}}, \bibinfo {author} {\bibfnamefont {X.}~\bibnamefont {Qiu}}, \bibinfo
  {author} {\bibfnamefont {X.}~\bibnamefont {Han}}, \bibinfo {author}
  {\bibfnamefont {Y.}~\bibnamefont {Peng}},\ and\ \bibinfo {author}
  {\bibfnamefont {X.}~\bibnamefont {Zhang}},\ }\href
  {https://doi.org/10.1063/1.5142562} {\bibfield  {journal} {\bibinfo
  {journal} {Appl. Phys. Lett.}\ }\textbf {\bibinfo {volume} {116}},\ \bibinfo
  {pages} {142404} (\bibinfo {year} {2020})}\BibitemShut {NoStop}%
\bibitem [{\citenamefont {Montoya}\ \emph {et~al.}(2017)\citenamefont
  {Montoya}, \citenamefont {Couture}, \citenamefont {Chess}, \citenamefont
  {Lee}, \citenamefont {Kent}, \citenamefont {Henze}, \citenamefont {Sinha},
  \citenamefont {Im}, \citenamefont {Kevan}, \citenamefont {Fischer},
  \citenamefont {McMorran}, \citenamefont {Lomakin}, \citenamefont {Roy},\ and\
  \citenamefont {Fullerton}}]{Montoya2017a}%
  \BibitemOpen
  \bibfield  {author} {\bibinfo {author} {\bibfnamefont {S.~A.}\ \bibnamefont
  {Montoya}}, \bibinfo {author} {\bibfnamefont {S.}~\bibnamefont {Couture}},
  \bibinfo {author} {\bibfnamefont {J.~J.}\ \bibnamefont {Chess}}, \bibinfo
  {author} {\bibfnamefont {J.~C.~T.}\ \bibnamefont {Lee}}, \bibinfo {author}
  {\bibfnamefont {N.}~\bibnamefont {Kent}}, \bibinfo {author} {\bibfnamefont
  {D.}~\bibnamefont {Henze}}, \bibinfo {author} {\bibfnamefont {S.~K.}\
  \bibnamefont {Sinha}}, \bibinfo {author} {\bibfnamefont {M.-Y.}\ \bibnamefont
  {Im}}, \bibinfo {author} {\bibfnamefont {S.~D.}\ \bibnamefont {Kevan}},
  \bibinfo {author} {\bibfnamefont {P.}~\bibnamefont {Fischer}}, \bibinfo
  {author} {\bibfnamefont {B.~J.}\ \bibnamefont {McMorran}}, \bibinfo {author}
  {\bibfnamefont {V.}~\bibnamefont {Lomakin}}, \bibinfo {author} {\bibfnamefont
  {S.}~\bibnamefont {Roy}},\ and\ \bibinfo {author} {\bibfnamefont {E.~E.}\
  \bibnamefont {Fullerton}},\ }\href
  {https://doi.org/10.1103/PhysRevB.95.024415} {\bibfield  {journal} {\bibinfo
  {journal} {Phys. Rev. B}\ }\textbf {\bibinfo {volume} {95}},\ \bibinfo
  {pages} {024415} (\bibinfo {year} {2017})}\BibitemShut {NoStop}%
\bibitem [{\citenamefont {Heigl}\ \emph {et~al.}(2021)\citenamefont {Heigl},
  \citenamefont {Koraltan}, \citenamefont {Va$\check{n}$atka}, \citenamefont
  {Kraft}, \citenamefont {Abert}, \citenamefont {Vogler}, \citenamefont
  {Semisalova}, \citenamefont {Che}, \citenamefont {Ullrich}, \citenamefont
  {Schmidt}, \citenamefont {Hintermayr}, \citenamefont {Grundler},
  \citenamefont {Farle}, \citenamefont {Urbánek}, \citenamefont {Suess},\ and\
  \citenamefont {Albrecht}}]{Heigl2021}%
  \BibitemOpen
  \bibfield  {author} {\bibinfo {author} {\bibfnamefont {M.}~\bibnamefont
  {Heigl}}, \bibinfo {author} {\bibfnamefont {S.}~\bibnamefont {Koraltan}},
  \bibinfo {author} {\bibfnamefont {M.}~\bibnamefont {Va$\check{n}$atka}},
  \bibinfo {author} {\bibfnamefont {R.}~\bibnamefont {Kraft}}, \bibinfo
  {author} {\bibfnamefont {C.}~\bibnamefont {Abert}}, \bibinfo {author}
  {\bibfnamefont {C.}~\bibnamefont {Vogler}}, \bibinfo {author} {\bibfnamefont
  {A.}~\bibnamefont {Semisalova}}, \bibinfo {author} {\bibfnamefont
  {P.}~\bibnamefont {Che}}, \bibinfo {author} {\bibfnamefont {A.}~\bibnamefont
  {Ullrich}}, \bibinfo {author} {\bibfnamefont {T.}~\bibnamefont {Schmidt}},
  \bibinfo {author} {\bibfnamefont {J.}~\bibnamefont {Hintermayr}}, \bibinfo
  {author} {\bibfnamefont {D.}~\bibnamefont {Grundler}}, \bibinfo {author}
  {\bibfnamefont {M.}~\bibnamefont {Farle}}, \bibinfo {author} {\bibfnamefont
  {M.}~\bibnamefont {Urbánek}}, \bibinfo {author} {\bibfnamefont
  {D.}~\bibnamefont {Suess}},\ and\ \bibinfo {author} {\bibfnamefont
  {M.}~\bibnamefont {Albrecht}},\ }\href
  {https://doi.org/10.1038/s41467-021-22600-7} {\bibfield  {journal} {\bibinfo
  {journal} {Nat Commun}\ }\textbf {\bibinfo {volume} {12}},\ \bibinfo {pages}
  {2611} (\bibinfo {year} {2021})}\BibitemShut {NoStop}%
\bibitem [{\citenamefont {Hassan}\ \emph {et~al.}(2024)\citenamefont {Hassan},
  \citenamefont {Koraltan}, \citenamefont {Ullrich}, \citenamefont {Bruckner},
  \citenamefont {Serha}, \citenamefont {Levchenko}, \citenamefont {Varvaro},
  \citenamefont {Kiselev}, \citenamefont {Heigl}, \citenamefont {Abert},
  \citenamefont {Suess},\ and\ \citenamefont {Albrecht}}]{Hassan2024}%
  \BibitemOpen
  \bibfield  {author} {\bibinfo {author} {\bibfnamefont {M.}~\bibnamefont
  {Hassan}}, \bibinfo {author} {\bibfnamefont {S.}~\bibnamefont {Koraltan}},
  \bibinfo {author} {\bibfnamefont {A.}~\bibnamefont {Ullrich}}, \bibinfo
  {author} {\bibfnamefont {F.}~\bibnamefont {Bruckner}}, \bibinfo {author}
  {\bibfnamefont {R.~O.}\ \bibnamefont {Serha}}, \bibinfo {author}
  {\bibfnamefont {K.~V.}\ \bibnamefont {Levchenko}}, \bibinfo {author}
  {\bibfnamefont {G.}~\bibnamefont {Varvaro}}, \bibinfo {author} {\bibfnamefont
  {N.~S.}\ \bibnamefont {Kiselev}}, \bibinfo {author} {\bibfnamefont
  {M.}~\bibnamefont {Heigl}}, \bibinfo {author} {\bibfnamefont
  {C.}~\bibnamefont {Abert}}, \bibinfo {author} {\bibfnamefont
  {D.}~\bibnamefont {Suess}},\ and\ \bibinfo {author} {\bibfnamefont
  {M.}~\bibnamefont {Albrecht}},\ }\href@noop {} {\bibfield  {journal}
  {\bibinfo  {journal} {Nature Physics}\ }\textbf {\bibinfo {volume} {20}},\
  \bibinfo {pages} {615} (\bibinfo {year} {2024})}\BibitemShut {NoStop}%
\bibitem [{\citenamefont {Chumak}\ \emph {et~al.}(2017)\citenamefont {Chumak},
  \citenamefont {Serga},\ and\ \citenamefont {Hillebrands}}]{Chumak2017}%
  \BibitemOpen
  \bibfield  {author} {\bibinfo {author} {\bibfnamefont {A.~V.}\ \bibnamefont
  {Chumak}}, \bibinfo {author} {\bibfnamefont {A.~A.}\ \bibnamefont {Serga}},\
  and\ \bibinfo {author} {\bibfnamefont {B.}~\bibnamefont {Hillebrands}},\
  }\href {https://doi.org/10.1088/1361-6463/aa6a65} {\bibfield  {journal}
  {\bibinfo  {journal} {Journal of Physics D: Applied Physics}\ }\textbf
  {\bibinfo {volume} {50}},\ \bibinfo {pages} {244001} (\bibinfo {year}
  {2017})}\BibitemShut {NoStop}%
\bibitem [{\citenamefont {Lonsky}\ and\ \citenamefont
  {Hoffmann}(2020)}]{Lonsky2020}%
  \BibitemOpen
  \bibfield  {author} {\bibinfo {author} {\bibfnamefont {M.}~\bibnamefont
  {Lonsky}}\ and\ \bibinfo {author} {\bibfnamefont {A.}~\bibnamefont
  {Hoffmann}},\ }\href {https://doi.org/10.1063/5.0027042} {\bibfield
  {journal} {\bibinfo  {journal} {APL Materials}\ }\textbf {\bibinfo {volume}
  {8}},\ \bibinfo {pages} {100903} (\bibinfo {year} {2020})}\BibitemShut
  {NoStop}%
\bibitem [{\citenamefont {Chen}\ and\ \citenamefont {Ma}(2021)}]{Zhedong2021}%
  \BibitemOpen
  \bibfield  {author} {\bibinfo {author} {\bibfnamefont {Z.}~\bibnamefont
  {Chen}}\ and\ \bibinfo {author} {\bibfnamefont {F.}~\bibnamefont {Ma}},\
  }\href {https://doi.org/10.1063/5.0061832} {\bibfield  {journal} {\bibinfo
  {journal} {Journal of Applied Physics}\ }\textbf {\bibinfo {volume} {130}},\
  \bibinfo {pages} {090901} (\bibinfo {year} {2021})}\BibitemShut {NoStop}%
\bibitem [{\citenamefont {Titze}\ \emph
  {et~al.}(2024{\natexlab{a}})\citenamefont {Titze}, \citenamefont {Koraltan},
  \citenamefont {Schmidt}, \citenamefont {Möller}, \citenamefont {Bruckner},
  \citenamefont {Abert}, \citenamefont {Suess}, \citenamefont {Ropers},
  \citenamefont {Steil}, \citenamefont {Albrecht},\ and\ \citenamefont
  {Mathias}}]{Titze2024a}%
  \BibitemOpen
  \bibfield  {author} {\bibinfo {author} {\bibfnamefont {T.}~\bibnamefont
  {Titze}}, \bibinfo {author} {\bibfnamefont {S.}~\bibnamefont {Koraltan}},
  \bibinfo {author} {\bibfnamefont {T.}~\bibnamefont {Schmidt}}, \bibinfo
  {author} {\bibfnamefont {M.}~\bibnamefont {Möller}}, \bibinfo {author}
  {\bibfnamefont {F.}~\bibnamefont {Bruckner}}, \bibinfo {author}
  {\bibfnamefont {C.}~\bibnamefont {Abert}}, \bibinfo {author} {\bibfnamefont
  {D.}~\bibnamefont {Suess}}, \bibinfo {author} {\bibfnamefont
  {C.}~\bibnamefont {Ropers}}, \bibinfo {author} {\bibfnamefont
  {D.}~\bibnamefont {Steil}}, \bibinfo {author} {\bibfnamefont
  {M.}~\bibnamefont {Albrecht}},\ and\ \bibinfo {author} {\bibfnamefont
  {S.}~\bibnamefont {Mathias}},\ }\href
  {https://doi.org/https://doi.org/10.1002/adfm.202313619} {\bibfield
  {journal} {\bibinfo  {journal} {Adv. Funct. Mater.}\ }\textbf {\bibinfo
  {volume} {34}},\ \bibinfo {pages} {2313619} (\bibinfo {year}
  {2024}{\natexlab{a}})}\BibitemShut {NoStop}%
\bibitem [{\citenamefont {Ogawa}\ \emph {et~al.}(2015)\citenamefont {Ogawa},
  \citenamefont {Seki},\ and\ \citenamefont {Tokura}}]{Ogawa2015}%
  \BibitemOpen
  \bibfield  {author} {\bibinfo {author} {\bibfnamefont {N.}~\bibnamefont
  {Ogawa}}, \bibinfo {author} {\bibfnamefont {S.}~\bibnamefont {Seki}},\ and\
  \bibinfo {author} {\bibfnamefont {Y.}~\bibnamefont {Tokura}},\ }\href
  {https://doi.org/10.1038/srep09552} {\bibfield  {journal} {\bibinfo
  {journal} {Scientific Reports}\ }\textbf {\bibinfo {volume} {5}},\ \bibinfo
  {pages} {9552} (\bibinfo {year} {2015})}\BibitemShut {NoStop}%
\bibitem [{\citenamefont {Padmanabhan}\ \emph {et~al.}(2019)\citenamefont
  {Padmanabhan}, \citenamefont {Sekiguchi}, \citenamefont {Versteeg},
  \citenamefont {Slivina}, \citenamefont {Tsurkan}, \citenamefont {Bord\'acs},
  \citenamefont {K\'ezsm\'arki},\ and\ \citenamefont {van
  Loosdrecht}}]{Padmanabhan2019}%
  \BibitemOpen
  \bibfield  {author} {\bibinfo {author} {\bibfnamefont {P.}~\bibnamefont
  {Padmanabhan}}, \bibinfo {author} {\bibfnamefont {F.}~\bibnamefont
  {Sekiguchi}}, \bibinfo {author} {\bibfnamefont {R.~B.}\ \bibnamefont
  {Versteeg}}, \bibinfo {author} {\bibfnamefont {E.}~\bibnamefont {Slivina}},
  \bibinfo {author} {\bibfnamefont {V.}~\bibnamefont {Tsurkan}}, \bibinfo
  {author} {\bibfnamefont {S.}~\bibnamefont {Bord\'acs}}, \bibinfo {author}
  {\bibfnamefont {I.}~\bibnamefont {K\'ezsm\'arki}},\ and\ \bibinfo {author}
  {\bibfnamefont {P.~H.~M.}\ \bibnamefont {van Loosdrecht}},\ }\href
  {https://doi.org/10.1103/PhysRevLett.122.107203} {\bibfield  {journal}
  {\bibinfo  {journal} {Phys. Rev. Lett.}\ }\textbf {\bibinfo {volume} {122}},\
  \bibinfo {pages} {107203} (\bibinfo {year} {2019})}\BibitemShut {NoStop}%
\bibitem [{\citenamefont {Kalin}\ \emph {et~al.}(2022)\citenamefont {Kalin},
  \citenamefont {Sievers}, \citenamefont {F\"user}, \citenamefont {Schumacher},
  \citenamefont {Bieler}, \citenamefont {Garc\'{\i}a-S\'anchez}, \citenamefont
  {Bauer},\ and\ \citenamefont {Pfleiderer}}]{Kalin2022}%
  \BibitemOpen
  \bibfield  {author} {\bibinfo {author} {\bibfnamefont {J.}~\bibnamefont
  {Kalin}}, \bibinfo {author} {\bibfnamefont {S.}~\bibnamefont {Sievers}},
  \bibinfo {author} {\bibfnamefont {H.}~\bibnamefont {F\"user}}, \bibinfo
  {author} {\bibfnamefont {H.~W.}\ \bibnamefont {Schumacher}}, \bibinfo
  {author} {\bibfnamefont {M.}~\bibnamefont {Bieler}}, \bibinfo {author}
  {\bibfnamefont {F.}~\bibnamefont {Garc\'{\i}a-S\'anchez}}, \bibinfo {author}
  {\bibfnamefont {A.}~\bibnamefont {Bauer}},\ and\ \bibinfo {author}
  {\bibfnamefont {C.}~\bibnamefont {Pfleiderer}},\ }\href
  {https://doi.org/10.1103/PhysRevB.106.054430} {\bibfield  {journal} {\bibinfo
   {journal} {Phys. Rev. B}\ }\textbf {\bibinfo {volume} {106}},\ \bibinfo
  {pages} {054430} (\bibinfo {year} {2022})}\BibitemShut {NoStop}%
\bibitem [{\citenamefont {Titze}\ \emph {et~al.}(2025)\citenamefont {Titze},
  \citenamefont {Koraltan}, \citenamefont {Matthies}, \citenamefont {Schmidt},
  \citenamefont {Suess}, \citenamefont {Albrecht}, \citenamefont {Mathias},\
  and\ \citenamefont {Steil}}]{Titze2025}%
  \BibitemOpen
  \bibfield  {author} {\bibinfo {author} {\bibfnamefont {T.}~\bibnamefont
  {Titze}}, \bibinfo {author} {\bibfnamefont {S.}~\bibnamefont {Koraltan}},
  \bibinfo {author} {\bibfnamefont {M.}~\bibnamefont {Matthies}}, \bibinfo
  {author} {\bibfnamefont {T.}~\bibnamefont {Schmidt}}, \bibinfo {author}
  {\bibfnamefont {D.}~\bibnamefont {Suess}}, \bibinfo {author} {\bibfnamefont
  {M.}~\bibnamefont {Albrecht}}, \bibinfo {author} {\bibfnamefont
  {S.}~\bibnamefont {Mathias}},\ and\ \bibinfo {author} {\bibfnamefont
  {D.}~\bibnamefont {Steil}},\ }\href {https://doi.org/10.1103/wgrf-cw1t}
  {\bibfield  {journal} {\bibinfo  {journal} {Phys. Rev. B}\ }\textbf {\bibinfo
  {volume} {112}},\ \bibinfo {pages} {064413} (\bibinfo {year}
  {2025})}\BibitemShut {NoStop}%
\bibitem [{\citenamefont {Titze}\ \emph
  {et~al.}(2024{\natexlab{b}})\citenamefont {Titze}, \citenamefont {Koraltan},
  \citenamefont {Schmidt}, \citenamefont {Suess}, \citenamefont {Albrecht},
  \citenamefont {Mathias},\ and\ \citenamefont {Steil}}]{Titze2024b}%
  \BibitemOpen
  \bibfield  {author} {\bibinfo {author} {\bibfnamefont {T.}~\bibnamefont
  {Titze}}, \bibinfo {author} {\bibfnamefont {S.}~\bibnamefont {Koraltan}},
  \bibinfo {author} {\bibfnamefont {T.}~\bibnamefont {Schmidt}}, \bibinfo
  {author} {\bibfnamefont {D.}~\bibnamefont {Suess}}, \bibinfo {author}
  {\bibfnamefont {M.}~\bibnamefont {Albrecht}}, \bibinfo {author}
  {\bibfnamefont {S.}~\bibnamefont {Mathias}},\ and\ \bibinfo {author}
  {\bibfnamefont {D.}~\bibnamefont {Steil}},\ }\href
  {https://doi.org/10.1103/PhysRevLett.133.156701} {\bibfield  {journal}
  {\bibinfo  {journal} {Phys. Rev. Lett.}\ }\textbf {\bibinfo {volume} {133}},\
  \bibinfo {pages} {156701} (\bibinfo {year} {2024}{\natexlab{b}})}\BibitemShut
  {NoStop}%
\bibitem [{\citenamefont {Jefremovas}\ \emph {et~al.}(2025)\citenamefont
  {Jefremovas}, \citenamefont {Leutner}, \citenamefont {Fischer}, \citenamefont
  {Marqués-Marchán}, \citenamefont {Winkler}, \citenamefont {Asenjo},
  \citenamefont {Sinova}, \citenamefont {Frömter},\ and\ \citenamefont
  {Kläui}}]{Jefremovas2025}%
  \BibitemOpen
  \bibfield  {author} {\bibinfo {author} {\bibfnamefont {E.~M.}\ \bibnamefont
  {Jefremovas}}, \bibinfo {author} {\bibfnamefont {K.}~\bibnamefont {Leutner}},
  \bibinfo {author} {\bibfnamefont {M.~G.}\ \bibnamefont {Fischer}}, \bibinfo
  {author} {\bibfnamefont {J.}~\bibnamefont {Marqués-Marchán}}, \bibinfo
  {author} {\bibfnamefont {T.~B.}\ \bibnamefont {Winkler}}, \bibinfo {author}
  {\bibfnamefont {A.}~\bibnamefont {Asenjo}}, \bibinfo {author} {\bibfnamefont
  {J.}~\bibnamefont {Sinova}}, \bibinfo {author} {\bibfnamefont
  {R.}~\bibnamefont {Frömter}},\ and\ \bibinfo {author} {\bibfnamefont
  {M.}~\bibnamefont {Kläui}},\ }\href
  {https://doi.org/10.1016/j.newton.2025.100036} {\bibfield  {journal}
  {\bibinfo  {journal} {Newton}\ }\textbf {\bibinfo {volume} {1}},\ \bibinfo
  {pages} {100036} (\bibinfo {year} {2025})}\BibitemShut {NoStop}%
\bibitem [{\citenamefont {Kosevich}\ \emph {et~al.}(1990)\citenamefont
  {Kosevich}, \citenamefont {Ivanov},\ and\ \citenamefont
  {Kovalev}}]{Kosevich1990}%
  \BibitemOpen
  \bibfield  {author} {\bibinfo {author} {\bibfnamefont {A.}~\bibnamefont
  {Kosevich}}, \bibinfo {author} {\bibfnamefont {B.}~\bibnamefont {Ivanov}},\
  and\ \bibinfo {author} {\bibfnamefont {A.}~\bibnamefont {Kovalev}},\ }\href
  {https://doi.org/https://doi.org/10.1016/0370-1573(90)90130-T} {\bibfield
  {journal} {\bibinfo  {journal} {Physics Reports}\ }\textbf {\bibinfo {volume}
  {194}},\ \bibinfo {pages} {117} (\bibinfo {year} {1990})}\BibitemShut
  {NoStop}%
\bibitem [{\citenamefont {Hierro-Rodriguez}\ \emph {et~al.}(2020)\citenamefont
  {Hierro-Rodriguez}, \citenamefont {Quirós}, \citenamefont {Sorrentino},
  \citenamefont {Alvarez-Prado}, \citenamefont {Martín}, \citenamefont
  {Alameda}, \citenamefont {McVitie}, \citenamefont {Pereiro}, \citenamefont
  {Vélez},\ and\ \citenamefont {Ferrer}}]{HierroRodriguez2020}%
  \BibitemOpen
  \bibfield  {author} {\bibinfo {author} {\bibfnamefont {A.}~\bibnamefont
  {Hierro-Rodriguez}}, \bibinfo {author} {\bibfnamefont {C.}~\bibnamefont
  {Quirós}}, \bibinfo {author} {\bibfnamefont {A.}~\bibnamefont {Sorrentino}},
  \bibinfo {author} {\bibfnamefont {L.~M.}\ \bibnamefont {Alvarez-Prado}},
  \bibinfo {author} {\bibfnamefont {J.~I.}\ \bibnamefont {Martín}}, \bibinfo
  {author} {\bibfnamefont {J.~M.}\ \bibnamefont {Alameda}}, \bibinfo {author}
  {\bibfnamefont {S.}~\bibnamefont {McVitie}}, \bibinfo {author} {\bibfnamefont
  {E.}~\bibnamefont {Pereiro}}, \bibinfo {author} {\bibfnamefont
  {M.}~\bibnamefont {Vélez}},\ and\ \bibinfo {author} {\bibfnamefont
  {S.}~\bibnamefont {Ferrer}},\ }\href
  {https://doi.org/10.1038/s41467-020-20119-x} {\bibfield  {journal} {\bibinfo
  {journal} {Nature Communications}\ }\textbf {\bibinfo {volume} {11}},\
  \bibinfo {pages} {6382} (\bibinfo {year} {2020})}\BibitemShut {NoStop}%
\bibitem [{\citenamefont {Tejo}\ \emph {et~al.}(2021)\citenamefont {Tejo},
  \citenamefont {Heredero}, \citenamefont {Chubykalo-Fesenko},\ and\
  \citenamefont {Guslienko}}]{Tejo2021}%
  \BibitemOpen
  \bibfield  {author} {\bibinfo {author} {\bibfnamefont {F.}~\bibnamefont
  {Tejo}}, \bibinfo {author} {\bibfnamefont {R.~H.}\ \bibnamefont {Heredero}},
  \bibinfo {author} {\bibfnamefont {O.}~\bibnamefont {Chubykalo-Fesenko}},\
  and\ \bibinfo {author} {\bibfnamefont {K.~Y.}\ \bibnamefont {Guslienko}},\
  }\href {https://doi.org/10.1038/s41598-021-01175-9} {\bibfield  {journal}
  {\bibinfo  {journal} {Scientific Reports}\ }\textbf {\bibinfo {volume}
  {11}},\ \bibinfo {pages} {21714} (\bibinfo {year} {2021})}\BibitemShut
  {NoStop}%
\bibitem [{\citenamefont {Göbel}\ \emph {et~al.}(2021)\citenamefont {Göbel},
  \citenamefont {Mertig},\ and\ \citenamefont {Tretiakov}}]{Goebel2021}%
  \BibitemOpen
  \bibfield  {author} {\bibinfo {author} {\bibfnamefont {B.}~\bibnamefont
  {Göbel}}, \bibinfo {author} {\bibfnamefont {I.}~\bibnamefont {Mertig}},\
  and\ \bibinfo {author} {\bibfnamefont {O.~A.}\ \bibnamefont {Tretiakov}},\
  }\href {https://doi.org/https://doi.org/10.1016/j.physrep.2020.10.001}
  {\bibfield  {journal} {\bibinfo  {journal} {Phys. Rep.}\ }\textbf {\bibinfo
  {volume} {895}},\ \bibinfo {pages} {1} (\bibinfo {year} {2021})}\BibitemShut
  {NoStop}%
\bibitem [{\citenamefont {Zhang}\ \emph {et~al.}(2018)\citenamefont {Zhang},
  \citenamefont {Zhang}, \citenamefont {Wen}, \citenamefont {Chudnovsky},\ and\
  \citenamefont {Zhang}}]{Zhang2018b}%
  \BibitemOpen
  \bibfield  {author} {\bibinfo {author} {\bibfnamefont {S.}~\bibnamefont
  {Zhang}}, \bibinfo {author} {\bibfnamefont {J.}~\bibnamefont {Zhang}},
  \bibinfo {author} {\bibfnamefont {Y.}~\bibnamefont {Wen}}, \bibinfo {author}
  {\bibfnamefont {E.~M.}\ \bibnamefont {Chudnovsky}},\ and\ \bibinfo {author}
  {\bibfnamefont {X.}~\bibnamefont {Zhang}},\ }\href
  {https://doi.org/10.1038/s42005-018-0040-5} {\bibfield  {journal} {\bibinfo
  {journal} {Commun. Phys.}\ }\textbf {\bibinfo {volume} {1}},\ \bibinfo
  {pages} {36} (\bibinfo {year} {2018})}\BibitemShut {NoStop}%
\bibitem [{Note1()}]{Note1}%
  \BibitemOpen
  \bibinfo {note} {Note that these changes are considerably smaller than those
  observed experimentally, which is related to the smaller changes in density
  in the simulation compared to the MFM.}\BibitemShut {Stop}%
\bibitem [{\citenamefont {Zhang}\ \emph {et~al.}(2015)\citenamefont {Zhang},
  \citenamefont {Zhao}, \citenamefont {Fangohr}, \citenamefont {Liu},
  \citenamefont {Xia}, \citenamefont {Xia},\ and\ \citenamefont
  {Morvan}}]{Zhang2015}%
  \BibitemOpen
  \bibfield  {author} {\bibinfo {author} {\bibfnamefont {X.}~\bibnamefont
  {Zhang}}, \bibinfo {author} {\bibfnamefont {G.~P.}\ \bibnamefont {Zhao}},
  \bibinfo {author} {\bibfnamefont {H.}~\bibnamefont {Fangohr}}, \bibinfo
  {author} {\bibfnamefont {J.~P.}\ \bibnamefont {Liu}}, \bibinfo {author}
  {\bibfnamefont {W.~X.}\ \bibnamefont {Xia}}, \bibinfo {author} {\bibfnamefont
  {J.}~\bibnamefont {Xia}},\ and\ \bibinfo {author} {\bibfnamefont {F.~J.}\
  \bibnamefont {Morvan}},\ }\href {https://doi.org/10.1038/srep07643}
  {\bibfield  {journal} {\bibinfo  {journal} {Sci. Rep.}\ }\textbf {\bibinfo
  {volume} {5}},\ \bibinfo {pages} {7643} (\bibinfo {year} {2015})}\BibitemShut
  {NoStop}%
\bibitem [{\citenamefont {Castro}\ \emph {et~al.}(2020)\citenamefont {Castro},
  \citenamefont {Mancilla-Almonacid}, \citenamefont {Valdivia},\ and\
  \citenamefont {Allende}}]{Castro2020}%
  \BibitemOpen
  \bibfield  {author} {\bibinfo {author} {\bibfnamefont {M.~A.}\ \bibnamefont
  {Castro}}, \bibinfo {author} {\bibfnamefont {D.}~\bibnamefont
  {Mancilla-Almonacid}}, \bibinfo {author} {\bibfnamefont {J.~A.}\ \bibnamefont
  {Valdivia}},\ and\ \bibinfo {author} {\bibfnamefont {S.}~\bibnamefont
  {Allende}},\ }\href {https://doi.org/10.1088/1361-648X/ab6aec} {\bibfield
  {journal} {\bibinfo  {journal} {J. Phys. Condens. Matter}\ }\textbf {\bibinfo
  {volume} {32}},\ \bibinfo {pages} {175801} (\bibinfo {year}
  {2020})}\BibitemShut {NoStop}%
\bibitem [{\citenamefont {Bruckner}\ \emph {et~al.}(2023)\citenamefont
  {Bruckner}, \citenamefont {Koraltan}, \citenamefont {Abert},\ and\
  \citenamefont {Suess}}]{Bruckner2023}%
  \BibitemOpen
  \bibfield  {author} {\bibinfo {author} {\bibfnamefont {F.}~\bibnamefont
  {Bruckner}}, \bibinfo {author} {\bibfnamefont {S.}~\bibnamefont {Koraltan}},
  \bibinfo {author} {\bibfnamefont {C.}~\bibnamefont {Abert}},\ and\ \bibinfo
  {author} {\bibfnamefont {D.}~\bibnamefont {Suess}},\ }\href@noop {}
  {\bibfield  {journal} {\bibinfo  {journal} {Scientific Reports}\ }\textbf
  {\bibinfo {volume} {13}},\ \bibinfo {pages} {12054} (\bibinfo {year}
  {2023})}\BibitemShut {NoStop}%
\end{thebibliography}%

\end{document}